\documentclass[12pt,preprint]{aastex}

\shorttitle{Plaut Field Abundances}
\shortauthors{Johnson et al.}

\begin{document}

\title{Alpha Enhancement and the Metallicity Distribution Function of
Plaut's Window}

\author{
Christian I. Johnson\altaffilmark{1,2,7},
R. Michael Rich\altaffilmark{1},  
Jon P. Fulbright\altaffilmark{3,4},
Elena Valenti\altaffilmark{5}, and
Andrew McWilliam\altaffilmark{6}
}

\altaffiltext{1}{Department of Physics and Astronomy, UCLA, 430 Portola Plaza,
Box 951547, Los Angeles, CA 90095-1547, USA; cijohnson@astro.ucla.edu; 
rmr@astro.ucla.edu}

\altaffiltext{2}{Department of Astronomy, Indiana University,
Swain West 319, 727 East Third Street, Bloomington, IN 47405--7105, USA}

\altaffiltext{3}{Department of Physics and Astronomy, Johns Hopkins University,
Baltimore, MD 21218, USA; jfulb@skysrv.pha.jhu.edu}

\altaffiltext{4}{Visiting astronomer, Cerro Tololo Inter--American
Observatory, National Optical Astronomy Observatory, which are operated by the
Association of Universities for Research in Astronomy, under contract with the
National Science Foundation.}

\altaffiltext{5}{European Southern Observatory, Karl Schwarzschild\--Stra\ss e 
2, D\--85748 Garching bei M\"{u}nchen, Germany; evalenti@eso.org}

\altaffiltext{6}{Observatories of the Carnegie Institution of Washington, 
Pasadena, CA, USA; andy@obs.carnegiescience.edu}

\altaffiltext{7}{National Science Foundation Astronomy and Astrophysics 
Postdoctoral Fellow}

\begin{abstract}

We present Fe, Si, and Ca abundances for 61 giants in Plaut's Window 
(l=--1$\degr$,b=--8.5$\degr$) and Fe abundances for an additional 31 giants
in a second, nearby field (l=0$\degr$,b=--8$\degr$) derived from high 
resolution (R$\approx$25,000) spectra obtained with the Blanco 4m telescope 
and Hydra multifiber spectrograph.  The median metallicity of red giant branch 
(RGB) stars in the Plaut field is $\sim$0.4 dex lower than those in Baade's 
Window, and confirms the presence of an iron abundance gradient along the bulge
minor axis.  The full metallicity range of our (biased) RGB sample spans 
--1.5$<$[Fe/H]$<$+0.3, which is similar to that found in other bulge fields. 
We also derive a photometric metallicity distribution function for RGB stars 
in the (l=--1$\degr$,b=--8.5$\degr$) field and find very good agreement with 
the spectroscopic metallicity distribution.  The radial velocity and dispersion
data for the bulge RGB stars are in agreement with previous results of the 
\emph{BRAVA} survey, and we find evidence for a decreasing velocity dispersion 
with increasing [Fe/H].  The [$\alpha$/Fe] enhancement in Plaut field stars is
nearly identical to that observed in Baade's window, and suggests that an 
[$\alpha$/Fe] gradient does not exist between b=--4$\degr$ and --8$\degr$.  
Additionally, a subset of our sample (23 stars) appear to be foreground red 
clump stars that are very metal--rich, exhibit small metallicity and radial 
velocity dispersions, and are enhanced in $\alpha$ elements.  While these 
stars likely belong to the Galactic inner disk population, they exhibit 
[$\alpha$/Fe] ratios that are enhanced above the thin and thick disk.

\end{abstract}


\section{INTRODUCTION}

The advent of multifiber spectroscopy has enabled large scale surveys of 
abundances and kinematics of stars in the Galactic bulge.  The \emph{Bulge
Radial Velocity Assay} (\emph{BRAVA}) has explored the kinematics of M giants 
over the inner kpc and found the dynamics to be consistent with a rapidly 
rotating N--body bar that leaves little room for a classical bulge 
component (Shen et al. 2010).  These observations would argue for a relatively 
simple picture.  However, the large scale metallicity survey by Zoccali et al.
(2008) finds an abundance gradient in the outer bulge.  Nominally, this would 
be inconsistent with a purely dynamical process such as the buckling of a 
massive disk, as proposed by Shen et al. (2010).  Furthermore, Babusiaux et 
al. (2010) find that the metal--rich population appears to be more 
concentrated toward the plane, and that these stars also exhibit a larger 
velocity dispersion.  Observations in multiple bulge fields will be required 
to sort out this complicated picture.

The distance, differential reddening, and complex populations make it difficult
to quantify the internal age dispersion of the bulge.  Recent studies by 
Zoccali et al. (2003), using a statistical disk subtraction method, and 
Clarkson et al. (2008), using the proper motion separation method of Kuijken \&
Rich (2002), argue that the bulge is $>$90$\%$ dominated by a stellar 
population comparable in age to the inner halo globular clusters.  A trace 
population of stars brighter than the old main sequence turnoff is interpreted 
to be a foreground population by Feltzing \& Gilmore (2000).  Subsequent 
studies confirm this to be the foreground disk based on proper motions 
(Kuijken \& Rich 2002) and based on subtraction of an equivalent foreground 
disk population (Zoccali et al. 2003).

The evidence for early, rapid formation is also supported by the detailed 
chemical abundances.  McWilliam \& Rich (1994), followed by subsequent
studies (Rich \& Origlia 2005; Cunha \& Smith 2006; Fulbright et al. 2007; 
Lecureur et al. 2007; Rich et al. 2007; Alves--Brito et al. 2010; Bensby et al.
2010), found evidence that multiple $\alpha$ elements are 
enhanced in bulge stars.  This is consistent with early, rapid enrichment from 
Type II supernovae (SNe; e.g., Ballero et al. 2007).  However, it is not clear 
whether the $\alpha$ enhancement is present only in the inner bulge, or is 
found over a larger volume.  Shen et al. (2010) argue that there is a kinematic
unity of the bulge extending from 300 to 1000 pc -- essentially the entire 
optical bulge.  Models of the deprojected photometry of the bulge are 
consistent with a peanut--shaped bulge dominated by a bar with its major axis 
aligned at $\sim$20$\degr$ from the Galactic Center.  

Although there is a kinematic unity to the bulge, the picture from abundances 
is complex.  The observed metallicity gradient requires that the enrichment 
history of the bulge varied as a function of location.  Extragalactic analogs 
of the Milky Way, like NGC 4565, also have well established metallicity 
gradients (e.g., Proctor et al. 2000).  NGC 4565 has a peanut--shaped 
pseudobulge that is rotationally supported like the Milky Way bulge.  To date, 
large sample studies have not addressed whether the $\alpha$ enhancement 
observed in Baade's Window (e.g., McWilliam \& Rich 1994) and inner bulge 
fields (Rich et al. 2007) is also characteristic across the entire bulge, 
including the outer bulge fields at $>$1 kpc.  However, we note that Lecureur
et al. (2007) analyzed 5 stars at b=--12$\degr$ ($\sim$1.7 kpc below the 
Galactic plane) and did not find a change in the abundance patterns of 
[O/Fe]\footnote{We make use of the standard spectroscopic
notation where [A/B]$\equiv$log(N$_{\rm A}$/N$_{\rm B}$)$_{\rm star}$--
log(N$_{\rm A}$/N$_{\rm B}$)$_{\sun}$ and
log $\epsilon$(A)$\equiv$log(N$_{\rm A}$/N$_{\rm H}$)+12.0 for elements A and
B.} or [Mg/Fe], compared to Baade's Window.

Here we report a new analysis of 92 stars toward Plaut's low extinction window
(l=0$\degr$, b=--8 $\degr$), obtained using the Hydra 
multifiber spectrograph on the CTIO Blanco 4m telescope, in the echelle mode.  
We find that this minor axis field located roughly 1 kpc from the nucleus shows
a marked decline in mean metallicity compared to the b=--4$\degr$ field of 
Baade's Window.  However, $\alpha$ elements are enhanced as they are in Baade's
Window; this argues for a common enrichment history of early, rapid formation 
over the full volume of the bulge extending to 1 kpc.  The metallicity decline 
is present at the same latitude where the bulge is demonstrated to show 
cylindrical rotation (Howard et al. 2009; Shen et al. 2010).

The Galactic bulge at b=--8$\degr$ is in a transition from the inner bulge 
into the halo.  Although Zoccali et al. (2008) shows that there is a modest 
metallicity gradient from b=--4$\degr$ to b=--6$\degr$, the same study finds 
dramatically lower abundances at b=--12$\degr$, but the possibility of disk 
contamination at this latitude is also greater.  It is clear that 
the b=--8$\degr$ field is an important missing link in defining the bulge 
metallicity gradient.  Additionally, the M giants in this field have been shown 
by the \emph{BRAVA} project to participate in the dynamics of the bar.  We 
recall also that Howard et al. (2008) find no evidence for kinematic 
subpopulations in the radial velocity (RV) distribution of M giants, in these 
fields.  The present dataset enables us to explore how dynamics depend on 
chemical composition, albeit for a more modest sample.

\section{SPECTROSCOPIC OBSERVATIONS AND REDUCTIONS}

The spectra for this project were obtained at Cerro Tololo Inter--American
Observatory (CTIO) using the Blanco 4m telescope and Hydra multifiber 
spectrograph.  The observations covered two separate runs spanning 2006 May
27--28 and 2007 May 19--23, and targeted the approximate spectral regions of 
6000--6250, 6150--6400, 6500--6800, and 7650--7950 \AA.  All spectrograph 
setups employed the ``large" 300 $\micron$ (2$\arcsec$) fibers, 400 mm Bench 
Schmidt camera, 316 line mm$^{\rm -1}$ Echelle grating, and 100 $\micron$ slit 
plate to achieve a resolving power of 
R($\lambda$/$\Delta$$\lambda$)$\approx$25,000.  These spectrograph setups allow
for abundance determinations of several light odd--Z, $\alpha$, Fe--peak, and 
neutron--capture elements.  However, the focus of this paper is on iron and 
the heavier $\alpha$ elements silicon and calcium.  The remaining elemental 
abundances will be published in a forthcoming paper.

Two separate fields were targeted in Plaut's low--extinction window near the 
bulge minor axis at (l,b)=(0$\degr$,--8$\degr$) and (--1$\degr$,--8.5$\degr$).  
Unfortunately, the 6100--6200 \AA\ region, which contains most of the useful 
silicon and calcium lines relevant here, was only observed for the 
(l=--1$\degr$,b--8.5$\degr$) field.  The optical photometry for this project
was obtained from observations at the Las Campanas Swope 40 inch telescope (see
$\S$3), and the infrared photometry was obtained from the Two Micron All Sky 
Survey Database (2MASS; Skrutskie et al. 2006).\footnote{The 2MASS catalog can 
be accessed online at: http://irsa.ipac.caltech.edu/applications/Gator/.}  The 
observing program targeted stars with V magnitudes between approximately 12.5 
and 15.0 (9.5 to 12.0 in K$_{\rm s}$), and a B--V color range from about 1.0 to
3.0 (0.6 to 1.1 in J--K$_{\rm s}$).  The observed stars for each field are 
shown in Figure \ref{f1}.  

Although we attempted to obtain an unbiased sample spanning the full color 
breadth of the bulge red giant branch (RGB), the final data set 
for each field unfortunately misses the reddest RGB stars.  However, the depth 
along the bulge minor axis line--of--sight provides some protection against a 
strong metallicity bias, and a simple calculation (see $\S$6.1) suggests our 
RGB sample is probably not too severely biased.  Note that we have also 
observed 23 stars, included in the total sample of 92 stars, along the 
vertical sequence blueward of the RGB (filled cyan circles in Figure \ref{f1}).
We suspect these are foreground red clump stars that may belong to the inner 
Galactic disk population (see also $\S$6).

All data reduction was carried out using standard tasks provided in 
IRAF.\footnote{IRAF is distributed by the National Optical Astronomy 
Observatory, which is operated by the Association of Universities for Research 
in Astronomy, Inc., under cooperative agreement with the National Science 
Foundation.}  Overscan trimming and bias subtraction were applied using the 
\emph{ccdproc} IRAF routine.  The majority of the raw data reduction, including
fiber tracing, scattered light removal, flat--field correction, wavelength
calibration with ThAr comparison spectra, cosmic ray removal, background sky 
subtraction, and object spectrum extraction, was performed via the 
\emph{dohydra} task. In all cases the reduced spectra were continuum 
flattened, corrected for telluric contamination, and combined.  The final, 
co--added spectra ranged in signal--to--noise (S/N) from $\sim$50--100 per 
pixel.  

\section{PHOTOMETRIC OBSERVATIONS AND REDUCTIONS}

A set of B, V and I images for both bulge fields analyzed here 
spectroscopically were obtained at the Las Campanas Observatory in August 
2002.  We used the optical imager mounted at the Swope 40 inch telescope and 
equipped with the SITe\#3 detector.  The detector is characterized by a 
0.435$\arcsec$ pixel size, which provides a total field of view of 
14.8$\arcmin {\times}$ 22.8$\arcmin$.  Typical exposure times for individual 
images were 240, 120, and 60 seconds in the B, V, and I bands, respectively.  
During the observations the average seeing was $\sim$1--1.5$\arcsec$ (FWHM).  

All of the raw frames have been bias and flat--field corrected by means of 
standard IRAF routines, using a set of sky flat--fields taken during the same 
night.  The PSF--fitting procedure was performed independently on each image 
using the ALLSTAR/DAOPHOTII package (Stetson 1987).  A reasonable estimate of 
the internal photometric accuracy 
($\sigma _B \sim \sigma _V \sim \sigma _I \sim$ 0.03 mag) has been obtained from
the frame--to--frame rms scatter of multiple star measurements.  Aperture 
photometry with the PHOT/DAOPHOTII routine was performed on a large sample of
isolated bright stars across each frame in order to correct the PSF 
single--band catalogs.  These have been corrected for exposure time and 
airmass.  In particular, the atmospheric extinction coefficients were directly
derived by repeated observation of Landolt (1992) standard fields at different
airmasses.  A final catalog listing the instrumental B, V, and I magnitudes was
generated by cross--correlating the single--band catalogs, and the absolute 
calibration was obtained from several repeat observations of the same standard 
fields, finding negligible color terms.  We estimate an overall uncertainty of
$\pm$0.05 mag in the zero--point calibration for all the three bands.

The optical catalog was then combined with the J, H, and K$_{\rm s}$ photometry
from 2MASS.  This was performed in order to transform the star positions onto
the 2MASS coordinate system, and provided rms residuals of $\sim$0.2$\arcsec$
in both right ascension and declination.

\section{SPECTROSCOPIC ANALYSIS}

\subsection{Model Stellar Atmospheres}

The data analysis process mostly followed the procedures outlined in 
Fulbright et al. (2006; 2007) and Johnson \& Pilachowski (2010).  To briefly
summarize, effective temperature (T$_{\rm eff}$) and surface gravity (log g)
estimates were calculated primarily from V and K$_{\rm s}$ band photometry,
and the metallicity ([Fe/H]) and microturbulence (V$_{\rm t}$) parameters were 
determined spectroscopically.  These values were used to generate suitable 
one--dimensional model atmospheres (without convective overshoot) through 
interpolation within the $\alpha$--rich\footnote{Although these models assume 
[$\alpha$/Fe]=$+$0.4 for \emph{all} $\alpha$ elements, we only have [Si/Fe] and
[Ca/Fe] abundances and therefore are unable to verify the validity of this 
approximation.  However, Fulbright et al. (2007) found that the effects of 
using an $\alpha$--rich versus solar composition model atmosphere for a 
\emph{relative} abundance analysis were small and result in an average increase
of $+$0.06$\pm$0.02 dex in [Fe/H], an average increase of $+$0.03$\pm$0.02 dex 
in [Si/Fe], and an average decrease in [Ca/Fe] of --0.02$\pm$0.02 dex.} ODFNEW 
ATLAS9 grid (Castelli et al. 1997).\footnote{The model atmosphere grid can be 
downloaded from http://wwwuser.oat.ts.astro.it/castelli/grids.html.}  
Specifically, effective temperatures were calculated from the V--K 
color--temperature relation in Alonso et al. (1999; 2001), and surface 
gravities were calculated through the standard relation,
\begin{equation}
log(g_{*})=0.40(M_{bol.}-M_{bol.\sun})+log(g_{\sun})+4[log(T/T_{\sun})]+
log(M/M_{\sun}),
\end{equation}
with an assumed stellar mass of 0.80 M$_{\rm \sun}$ and distance of 8 kpc.
The model atmosphere metallicity values were initially set at [Fe/H]=--0.3, and
then iteratively adjusted to match the derived [Fe/H] ratio from the equivalent
width (EW) analysis.  Similarly, the microturbulent velocity was set at 2
km s$^{\rm -1}$ for all stars, and then improved following the method outlined 
in Magain (1984) that removes trends in Fe I abundance with line strength.

Although the average color excess in Plaut's field is E(B--V)$\approx$0.2, 
differential reddening exists at about the 15$\%$ level across Hydra's 
40$\arcmin$ field.  Therefore, target stars were individually dereddened using 
the NASA/IPAC Extragalactic Database Galactic Extinction 
Calculator\footnote{The extinction calculator can be accessed at: 
http://nedwww.ipac.caltech.edu/forms/calculator.html.}, which is based on the 
Schlegel et al. (1998) infrared maps.

\begin{figure}
\epsscale{0.80}
\plotone{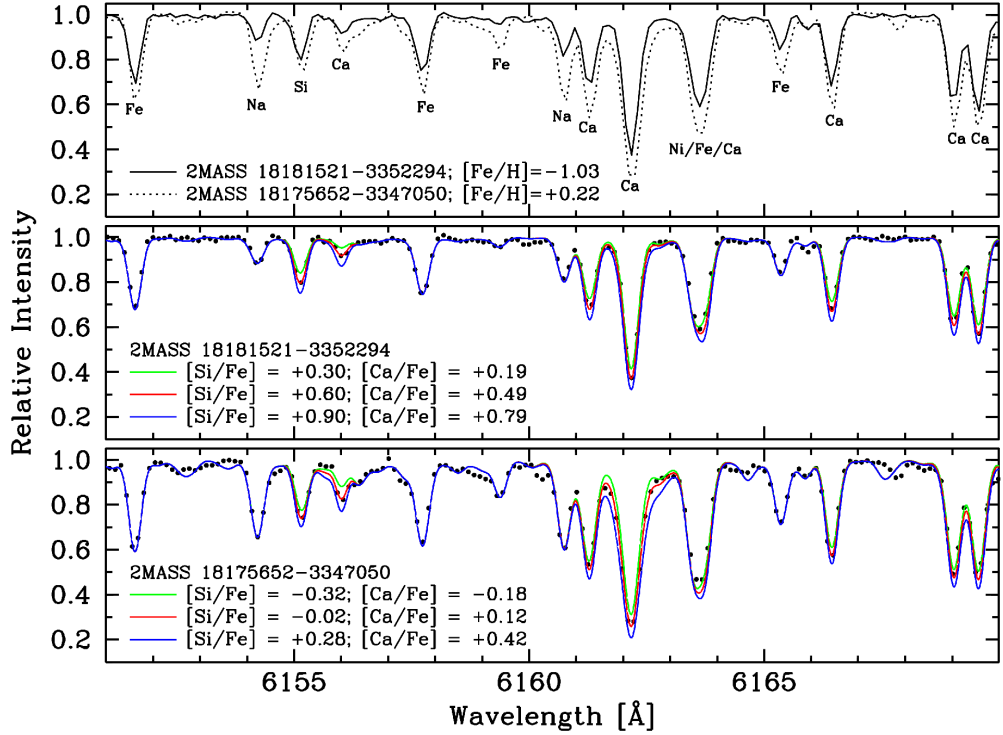}
\caption{The top panel illustrates the contrasting line strengths between two
stars in our bulge RGB sample with similar effective temperatures
(T$_{\rm eff}$$\approx$4200 K) but different metallicities.  Several
key lines are identified for guidance.  The middle and bottom panels show
sample synthetic spectrum fits to the data with the [Si/Fe] and [Ca/Fe] ratios
set at the best--fit values (red lines), and also altered by --0.30 dex (green
lines) and +0.30 dex (blue lines).}
\label{f2}
\end{figure}

\begin{figure}
\epsscale{0.80}
\plotone{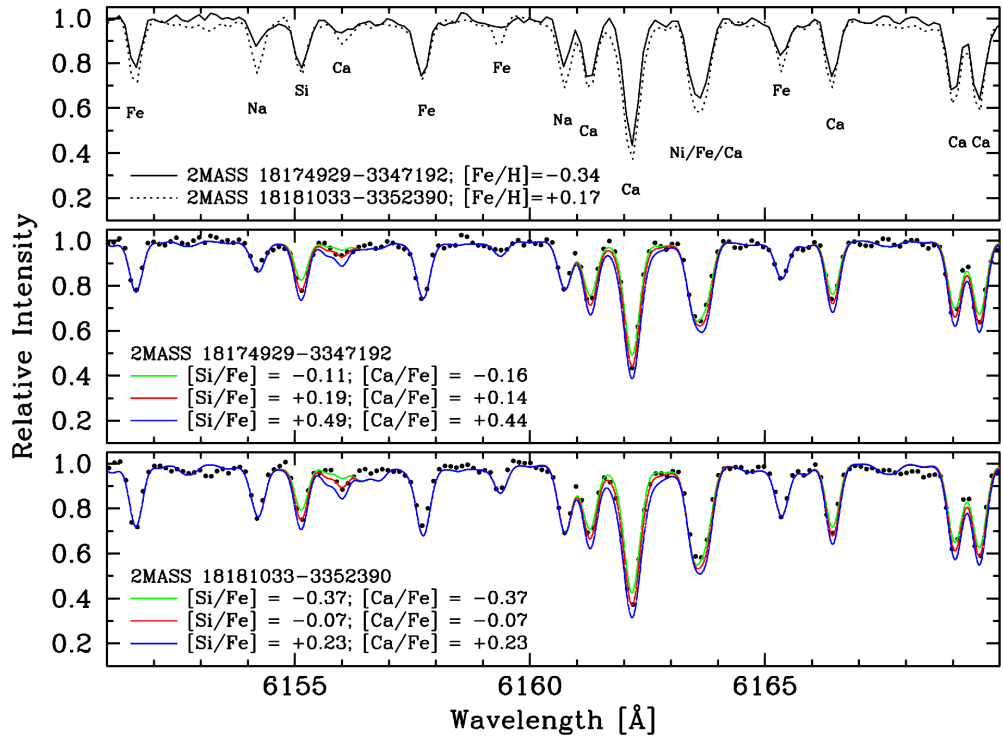}
\caption{The lines and symbols are the same as in Figure \ref{f2}, but the 
spectra shown here are from the foreground red clump sample 
(T$_{\rm eff}$$\approx$4900 K).}
\label{f3}
\end{figure}

Interestingly, the photometric surface gravity estimates for stars populating 
the vertical blue sequence seen in Figure \ref{f1} appeared too low compared 
to the cooler stars clearly populating the RGB.  This surface gravity 
discrepancy suggests either that the bluer stars are closer to the Sun than 8
kpc or that a problem exists with the bolometric magnitude calculation.  
Therefore, we used equation 1 to compare 
the surface gravity values obtained when using the V magnitude bolometric 
correction from Alonso et al. (1999) and the K magnitude bolometric correction 
from Buzzoni et al. (2010).  Although the bolometric magnitudes calculated 
from the K band will be less sensitive to reddening uncertainties, we found 
that the difference between the two surface gravity estimates was $<$0.05 dex. 
We interpret this result as evidence supporting the idea that the bluest stars 
in our sample are foreground red clump stars located $\sim$2--4 kpc from the 
Sun\footnote{Note that the distance estimate comes from comparing the 
photometric surface gravities with the theoretical gravities provided by
the Padova evolutionary tracks.  Note also that the few designated RGB stars 
lying blueward of the color cut--off in Figure \ref{f1} were relatively
metal--poor, and a comparison between their photometric and ``expected" surface
gravities, based on the Padova tracks, suggested they were at a distance of 
$\sim$8 kpc.}.  For these stars we adopted appropriate log g 
values from the Padova stellar evolutionary tracks (Girardi et al. 2000), and 
also used the pressure sensitive 6162 \AA\ Ca I line as a secondary guide 
(e.g., see Figures \ref{f2}--\ref{f3}).  Final model atmosphere parameters for 
all stars are provided in Table 1.

\subsection{Abundance Determinations}

The abundance measurement procedures for both the EW and spectrum synthesis 
analyses generally followed the methods outlined in Fulbright et al. (2006; 
2007) and Johnson \& Pilachowski (2010).  Specifically, Fe I abundances
were derived from a standard EW analysis where line profiles were fit with 
either a single Gaussian or deblended using multiple Gaussians via the 
interactive EW fitting code developed for Johnson et al. (2008).  The Fe I 
line list was taken from Fulbright et al. (2006) in the available wavelength 
windows listed in $\S$2 ($\sim$25--30 lines on average).  However, the 
individual log gf values were redetermined by forcing the EWs measured in the 
Arcturus atlas (Hinkle et al. 2000\footnote{The Arcturus atlas can be 
downloaded at: http://www.noao.edu/archives.html.}) to match the [Fe/H]=--0.51
ratio derived by Fulbright et al. (2006), using the same Arcturus atmospheric 
parameters and $\alpha$--rich ODFNEW ATLAS9 model atmosphere.  The final [Fe/H]
abundances listed in Table 1 were determined using the derived EWs, 
Arcturus--based line list, and the \emph{abfind} driver in the LTE line 
analysis code MOOG (Sneden 1973).

The neutral Si and Ca abundances reported here were derived from full spectrum 
synthesis of the 6000--6250 \AA\ region using the \emph{synth} driver in MOOG.
Sample synthetic spectrum fits to a few Si and Ca lines in metal--poor and
metal--rich stars of similar T$_{\rm eff}$ are shown in Figure \ref{f2} for 
the bulge RGB sample and Figure \ref{f3} for the foreground red clump stars.  
Although a few additional Si and Ca lines are available in the other wavelength
regions listed in $\S$2, the 6000--6250 \AA\ region spectra had significantly 
higher S/N.  Unfortunately, only ``Field 1" listed in Table 1 was observed in 
this wavelength setup, and therefore we only provide [Si/Fe] and [Ca/Fe] 
abundances for about two--thirds of the total sample.  

The log gf values for all Si and Ca lines were determined by fitting the 
synthetic spectra to the Arcturus spectrum and forcing [Si/Fe]=+0.35 and 
[Ca/Fe]=+0.21 (Fulbright et al. 2007).  Additionally, the log gf values of 
nearby metal lines were adjusted to match the Arcturus spectrum, using the 
abundances provided in Fulbright et al. (2007).  Since C and N abundances 
were not published in Fulbright et al. (2007), we used the [C/Fe] and [N/Fe] 
ratios from Peterson et al. (1993) in order to set the log gf values for CN 
lines.

As a consistency check, we tested the effects of normalizing the log gf values 
to the solar spectrum and the metal--rich ([Fe/H]$\approx$+0.35; Gratton \& 
Sneden 1990) giant $\mu$ Leo, which was obtained from the ELODIE archive 
(Moultaka et al. 2004)\footnote{Based on spectral data retrieved from the 
ELODIE archive at Observatoire de Haute--Provence (OHP).}.  When comparing the 
abundances derived from the Arcturus based log gf values to those derived from 
the solar based log gf values, we found a typical systematic offsets of 
$\sim$0.05, 0.05, and 0.10 dex for log $\epsilon$(Fe), log $\epsilon$(Si), and 
log $\epsilon$(Ca), respectively.  Similarly, the systematic offsets when 
comparing the Arcturus and $\mu$ Leo based log gf scales were approximately
0.03, 0.05, and 0.10 dex for log $\epsilon(Fe)$, log $\epsilon$(Si), and 
log $\epsilon$(Ca), respectively.  In both cases the [Ca/Fe] ratios experienced
larger changes than [Si/Fe] when transforming among the three abundance 
scales, presumably because of the significantly larger EWs of the available 
Ca I lines.

\subsection{Abundance Error Estimates}

In Table 2 we provide the calculated random ($\sigma$$_{\rm Rand.}$) and 
systematic ($\sigma$$_{\rm Sys.}$) errors for all elements analyzed in each
star.  The random error is defined here as the line--to--line dispersion in 
the derived log $\epsilon$(X) values for each element.  As mentioned above, 
the log $\epsilon$(Fe) abundances were typically based on $\sim$30 Fe I lines, 
and the log $\epsilon$(Si) and log $\epsilon$(Ca) abundances were based on
$\sim$3--5 lines each.  The average $\sigma$$_{\rm Rand.}$ values 
for Fe, Si, and Ca are 0.16$\pm$0.05, 0.07$\pm$0.05, and 0.09$\pm$0.05 dex, 
respectively.

The systematic errors were calculated following the procedure outlined in 
Fulbright et al. (2007; see also McWilliam et al. 1995), and were determined 
using an estimated uncertainty of T$_{\rm eff}$$\pm$50 K, log g$\pm$0.1 cgs, 
[M/H]$\pm$0.16 dex, and V$_{\rm t}$$\pm$0.3 km s$^{\rm -1}$.  The T$_{\rm eff}$
uncertainty represents the average change in our derived photometric 
temperatures if we assume a change in E(B--V) of $\pm$15$\%$, which is equal 
to the E(B--V) dispersion across our Hydra fields.  The surface gravity 
uncertainty represents the approximate change in the derived log (g) values if 
the assumed distance changes by +/-1 kpc.  The model [M/H] uncertainty is
taken from the average line--to--line dispersion in [Fe/H], and
the microturbulence uncertainty was chosen by examining the
spread in derived V$_{\rm t}$ values for stars of similar
T$_{\rm eff}$, log g, and [Fe/H].  The average
$\sigma$$_{\rm Sys.}$ values for Fe, Si, and Ca are 0.08$\pm$0.02,
0.09$\pm$0.02, and 0.08$\pm$0.01, respectively.  Note that the systematic 
uncertainty listed in Table 2 does not include the effects of normalizing the 
log gf values to either the solar or $\mu$ Leo abundance scales.

A secondary issue that could affect the foreground red clump stars is whether
the reddening values derived from the Schlegel et al. (1998) map are 
appropriate.  It is possible that the estimated reddening for these stars 
could be systematically too large, which would lead to photometric 
T$_{\rm eff}$ estimates, and consequently log g and [Fe/H] values, that are too
high.  Kunder et al. (2008) estimated an average E(B--V)$\approx$0.2 in Plaut's
window from observations of bulge RR Lyrae stars, and they found a full range in
color excess that spans approximately E(B--V)=0.1--0.3.  If we take E(B--V)=0.1
as a reasonable estimate of the \emph{minimum} color excess along the Plaut
field line--of--sight and assign this value to each clump star then the
average corresponding change to the photometric T$_{\rm eff}$ estimate is
about --150 K ($\sigma$=50 K).  

\begin{figure}
\epsscale{1.00}
\plotone{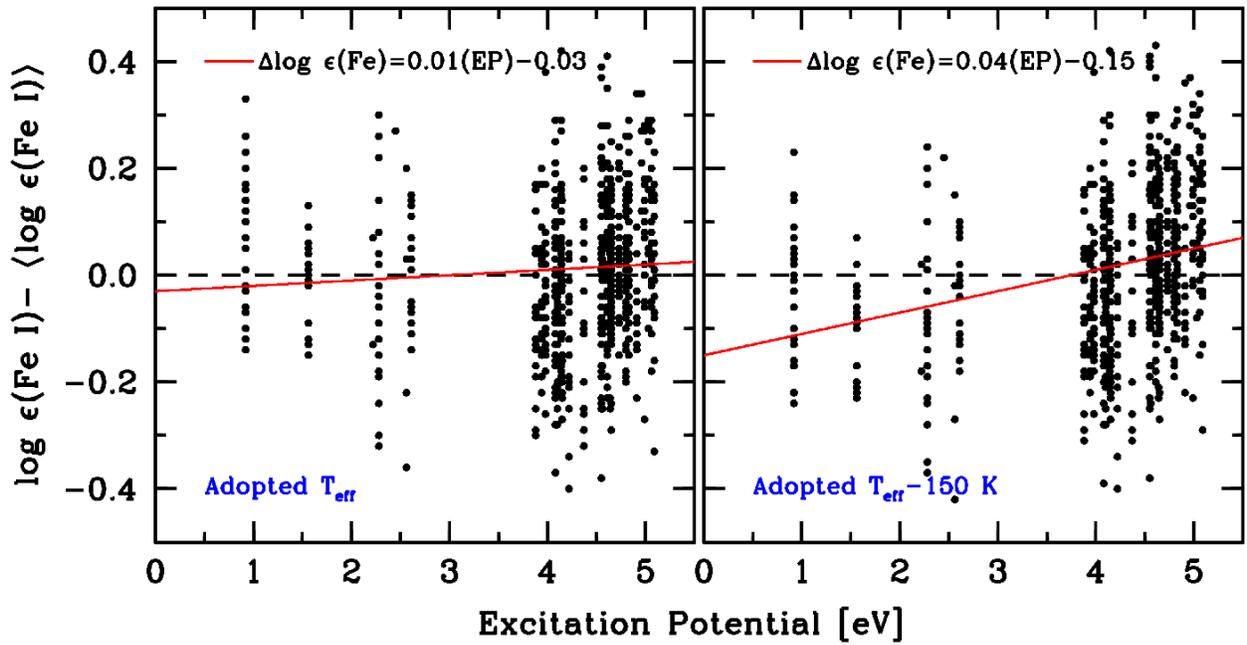}
\caption{The left panel shows 
log $\epsilon$(Fe I)--$\langle$log $\epsilon$(Fe I)$\rangle$ versus excitation
potential for all lines used in the red clump stars and for our adopted
T$_{\rm eff}$ values.  The right panel shows how the trend changes when the
temperatures are systematically lowered by 150 K.  In both panels the solid
red line indicates the least--squares fit to the data.}
\label{f4}
\end{figure}

In Figure \ref{f4} we show plots of 
log $\epsilon$(Fe)--$\langle$log $\epsilon$(Fe)$\rangle$ versus excitation
potential for all lines measured in all of the red clump stars, using our 
adopted T$_{\rm eff}$ values and a scale adjusted by --150 K.  Ideally, plots
such as Figure \ref{f4} should exhibit no slope if the adopted temperatures 
are correct, and the trend seen when using our adopted temperature scale 
indicates the red clump T$_{\rm eff}$ values employed here are suitable.  It is 
clear that systematically lowering the temperatures by 150 K produces a
significantly non--zero slope, and thus supports our claim that the Schlegel
et al. (1998) reddening values are an appropriate choice for the red clump 
stars.  Note that we can also reject the idea that the red clump stars are 
significantly less reddened from simple geometry.  We estimate these stars to 
be located $\sim$2--4 kpc from the Sun and thus $\sim$300--600 pc below the 
Galactic plane.  However, it is likely that most of the line--of--sight dust is
contained within roughly $\pm$100 pc of the Galactic plane (e.g., Marshall et
al. 2008).  This suggests that the majority of the line--of--sight reddening 
occurs in front of the red clump stars, and therefore both the bulge and 
foreground red clump stars should experience similar amounts of reddening.

\subsection{Radial Velocity Determinations}

Radial velocity measurements were performed using the IRAF task 
\emph{fxcor} to calculate the Fourier cross--correlation and \emph{rvcor} to 
calculate the heliocentric correction.  All program stars were measured against
the same high resolution, high S/N Arcturus atlas used in the abundance 
analysis, but the Arcturus template spectrum was smoothed and rebinned to match
Hydra's resolution.  The radial velocities were determined from the 
6150--6400 \AA\ window because it contains a large number of absorption lines 
and was observed for all target stars.  The final radial velocities for each
star are listed in Table 1, and typical measurement uncertainties reported by
\emph{fxcor} are $\sim$1 km s$^{\rm -1}$.

\section{PHOTOMETRIC ABUNDANCE ANALYSIS}

Although we obtained optical photometry for both bulge fields, here we only
provide a photometric abundance analysis for a field centered at 
(l,b)=(--0.89$\degr$,--8.45$\degr$), which includes stars in the 
spectroscopic field centered at (l,b)=(--1$\degr$,--8.5$\degr$).  In order to 
derive a metallicity distribution function for this 
field we have adopted an approach similar to that described in Bellazzini et 
al. (2003).  We compared the observed color--magnitude diagram with the 
empirical grid of cluster RGB ridge lines in the [M$_k$, (V\--K)$_0$] plane, 
which were carefully selected from the sample of Valenti et al. (2004a) to 
cover a wide metallicity range.  A photometric metallicity estimate for each 
star is then obtained from its color by interpolating within the grid of RGB 
templates.

\begin{figure}
\epsscale{1.00}
\plotone{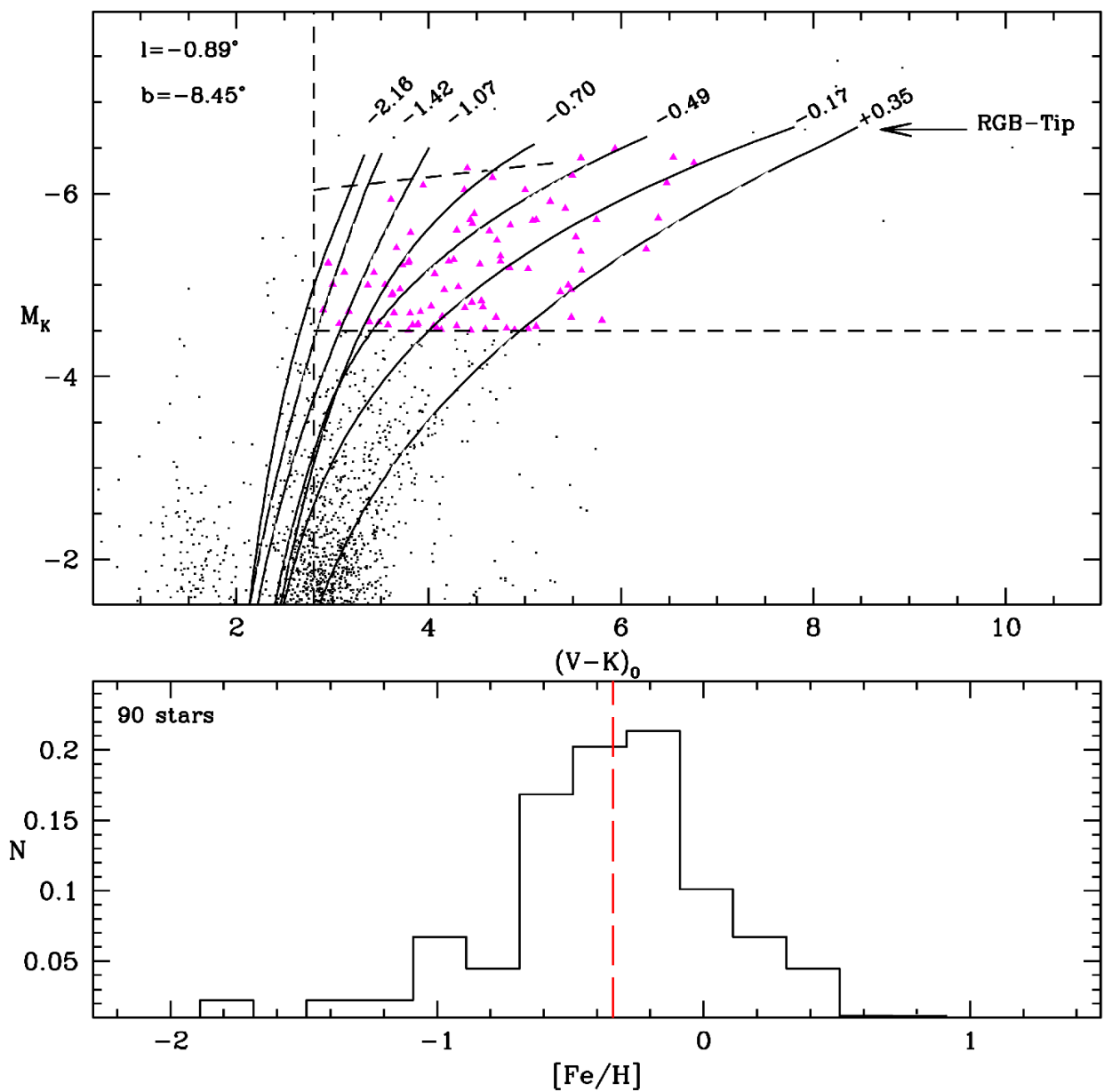}
\caption{The top panel shows the color--magnitude diagram of the bulge field
located at (l,b)=(--0.89$^{\degr}$,--8.45$^{\degr}$) in the absolute plane,
with the empirical RGB templates overplotted.  The filled magenta triangles
indicate the stars inside the dashed box used to derive the photometric
metallicity distribution function.  The bottom panel shows the derived
metallicity distribution function with a bin size of 0.2 dex, and the dashed
red line designates the median [Fe/H]=--0.34.}
\label{f5}
\end{figure}

The top panel of Figure \ref{f5} shows the result of the transformation of 
the observed color--magnitude diagram into the absolute plane.  We adopted a 
distance modulus of (m--M)$_0$=14.47, as measured by McNamara et al. (2000) 
using the RR Lyrae and $\delta$ Scuti variables in the Optical Gravitational 
Lensing Experiment (OGLE) survey, and a reddening E(B--V)=0.20, obtained by
averaging the most recent extinction estimates across the Bulge (e.g., 
Schlegel et al. 1998; Popowski et al. 2003).  The top panel of Figure \ref{f5}
also includes the grid of RGB fiducial ridge lines adopted from Valenti et al.
(2004a), along with the corresponding metallicity.  For empirical templates we
selected the Galactic globular clusters M 92, NGC 6752, NGC 288, 47 Tuc, NGC
6440, NGC 6528, and the old open cluster NGC 6791, in order to cover the
widest metallicity range with suitably fine steps.  In particular, NGC 6791 is
currently believed to be one of the most massive ($>$4000 M$_{\sun}$; Kaluzny 
\& Uldaski 1992), metal--rich ([Fe/H]$\sim$+0.35; Origlia et al. 2006), and 
oldest ($\sim$6--12 Gyr; Kaluzny \& Uldaski 1992; Carney et al. 2005) open 
clusters in the Galaxy.  This makes NGC 6791 an ideal template for the 
super--solar metallicity regime, and permitted full coverage from 
--2.16$\leq$[Fe/H]$\leq$+0.35 while avoiding major extrapolation beyond solar 
metallicity.

Unfortunately, the properties of the final metallicity distribution derived 
from similar procedures to the one used here can be sensitive to the choice of 
the template RGB grid used in the analysis.  All previous studies deriving 
metallicity distribution functions from RGB colors have employed their own 
custom--made recipe and reference grid.  Some authors adopt purely empirical, 
or alternatively purely theoretical grids, while others have adopted a mixture 
of both, using globular cluster data to set the zero--point of a given set of 
theoretical models or adding suitable calibrated isochrones to empirical 
templates in order to cover the missing metallicity range.  Although the 
main results are probably independent of these choices, there is little doubt
that the shape and other properties of the derived metallicity distribution 
functions are strongly affected by the details of the analysis, especially in
the high metallicity regime.  Here we use a pure {\it empirical} approach, 
and the accuracy of our results depend mainly on observable quantities (e.g.,
photometric and spectroscopic data; reddening estimates), whose errors can
be understood and minimized to some extent.  The use of theoretical models
would be easier and more precise in principle, as the grid can be uniformly
sampled.  However, theoretical models still lack a proper calibration with 
suitable empirical templates at super--solar metallicities.

In this work the metallicity distribution function has been computed for
stars within the dashed box shown in the top panel of Figure \ref{f5}.  We
determined the luminosity and color limits primarily to avoid contamination,
but also note that the RGB--tip region is more sensitive to [Fe/H] differences
than the lower RGB.  A lower luminosity limit was set at $M_K\leq4.5$ in order 
to retain the region of the RGB with the highest sensitivity to metallicity 
variations and to avoid contamination by red clump stars.  This choice also 
avoids the inclusion of asymptotic giant branch (AGB) clump stars, which are 
predicted to lie at $M_K\sim$--2.7 (Salasnish et al. 2000; Pietrinferni et al. 
2004).  In this way, the most populated region of the AGB is excluded and we 
can be confident that only a marginal fraction of AGB stars may contaminate our
sample.  The upper luminosity limit was set at the RGB--tip, as derived by 
Valenti et al. (2004b), to avoid the inclusion of possible bright AGB stars 
belonging to younger populations.  Finally, the blue color limit was set at 
(V--K)$_{\rm 0}$$\geq$2.8 to further minimize contamination from AGB and 
younger stars.  The final metallicity distribution function for the observed 
field, based on 90 stars, is shown in the bottom panel of Figure \ref{f5}.

\section{RESULTS AND DISCUSSION}

\subsection{Spectroscopic Metallicity and Radial Velocity Distributions}

\begin{figure}
\epsscale{1.00}
\plotone{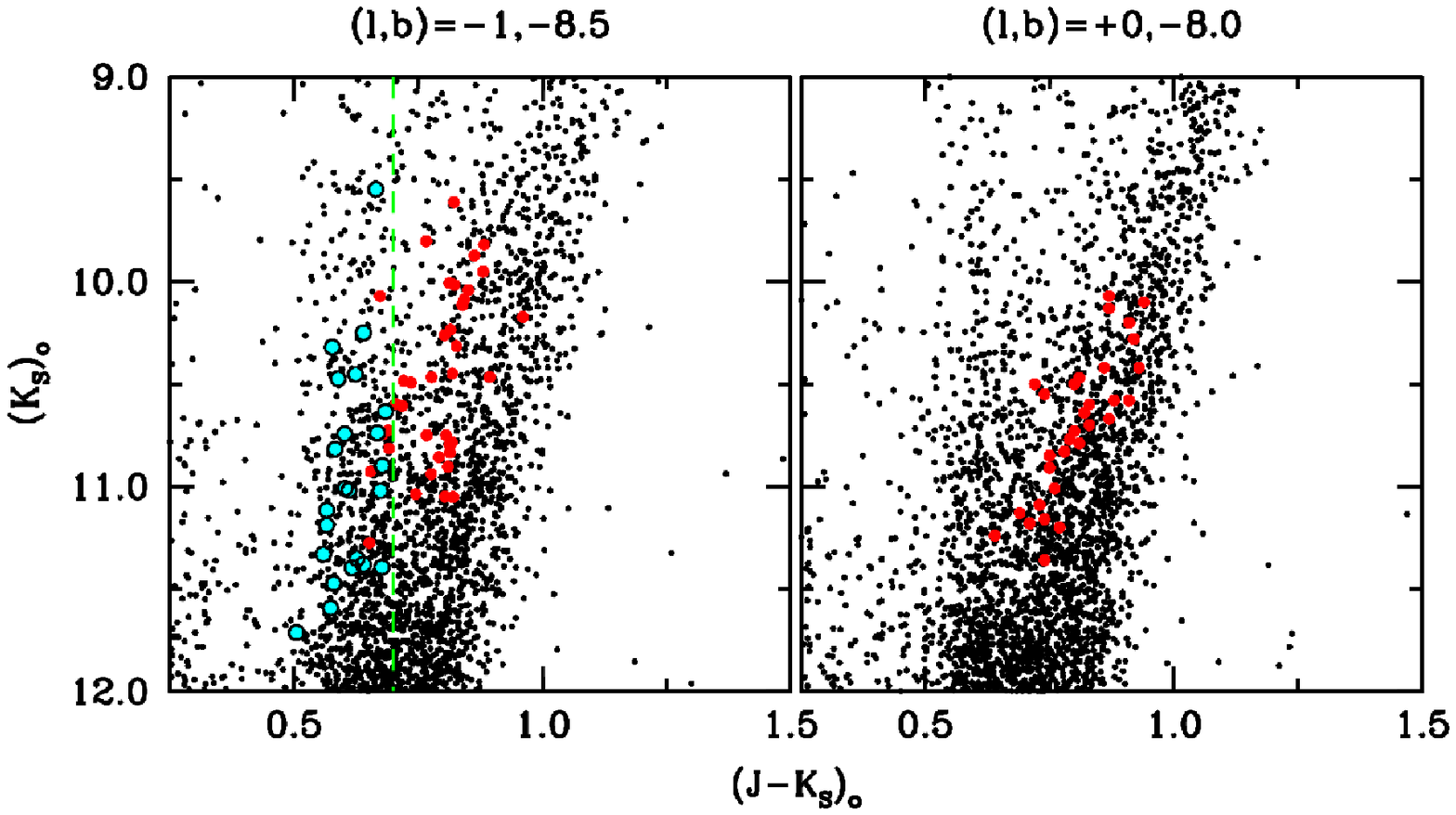}
\caption{Color--magnitude diagram of Plaut's window taken from the 2MASS 
database (filled black circles).  The two panels indicate the stars chosen for 
observation in this program.  The filled red circles in both panels indicate
probable RGB stars belonging to the bulge population, and the filled cyan
circles indicate probable foreground red clump stars.  In the left panel the 
dashed green line indicates the approximate color cutoff used to separate the 
RGB and clump populations (see text for details).}
\label{f1}
\end{figure}

\begin{figure}
\epsscale{1.00}
\plotone{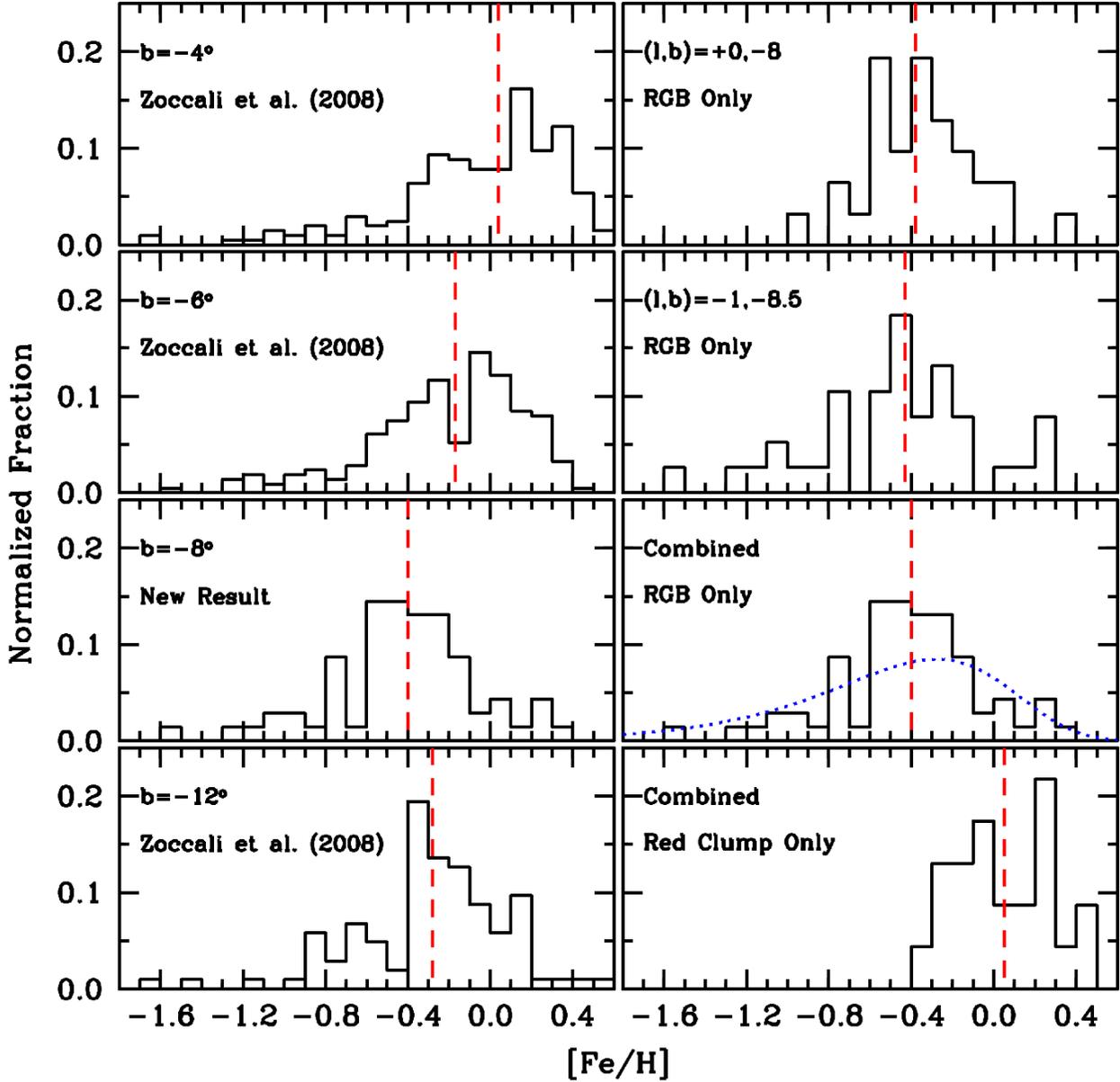}
\caption{Spectroscopic metallicity distribution functions of multiple
bulge fields in 0.1 dex bins.  For all panels the dashed red line designates
the median [Fe/H] value.  The left panels compare the metallicity distribution
functions of Zoccali et al. (2008) to our combined spectroscopic data at 
b=--8$\degr$ and b=--8.5$\degr$.  The right panels show the spectroscopic 
metallicity distribution functions for our two fields.  The dotted blue line 
shows the result of a one--zone, simple model calculation with a yield of 
z=0.0105.  Note that the area under the model curve has been scaled to equal 
the area under the data histogram.}
\label{f6}
\end{figure}

Although the bulge has long been known to exhibit a broad metallicity
range (e.g., Nassau \& Blanco 1958), the Zoccali et al. (2008) study was the 
first to quantify the shape of the metallicity distribution function at 
multiple Galactic latitudes with high resolution spectroscopy.  Their analysis 
showed that, at least along the minor axis beyond b=--4$\degr$, a radial 
metallicity gradient was present such that the average composition decreased 
from [Fe/H]$\approx$0 at b=--4$\degr$ to [Fe/H]$\approx$--0.3 at b=--12$\degr$.
Figure \ref{f6} shows our derived metallicity distribution function
in relation to the Zoccali et al. (2008) sample.  The new Plaut field data
confirm the existence of a minor axis metallicity gradient.  We find that 
the median metallicity of all stars (including RGB and clump) at b=--8$\degr$ 
is approximately a factor of two lower ([Fe/H]=--0.28) than the median 
metallicity of Baade's window (b=--4$\degr$; [Fe/H]=+0.04), and is also $>$0.1 
dex lower than the median metallicity at b=--6$\degr$ ([Fe/H]=--0.17).  The 
discrepancy is slightly larger if only the stars lying along the bulge RGB in 
Figure \ref{f1} are included.  This decreases the median metallicity in Plaut's
field to [Fe/H]=--0.40.  Interestingly, the median metallicity at b=--12$\degr$ 
([Fe/H]=--0.28) is essentially the same as, or even slightly higher than, the 
Plaut field metallicity.  This may suggest that the composition gradient 
levels off or becomes more shallow along the minor axis at distances $\ga$1 kpc
from the Galactic center.

\begin{figure}
\epsscale{1.00}
\plotone{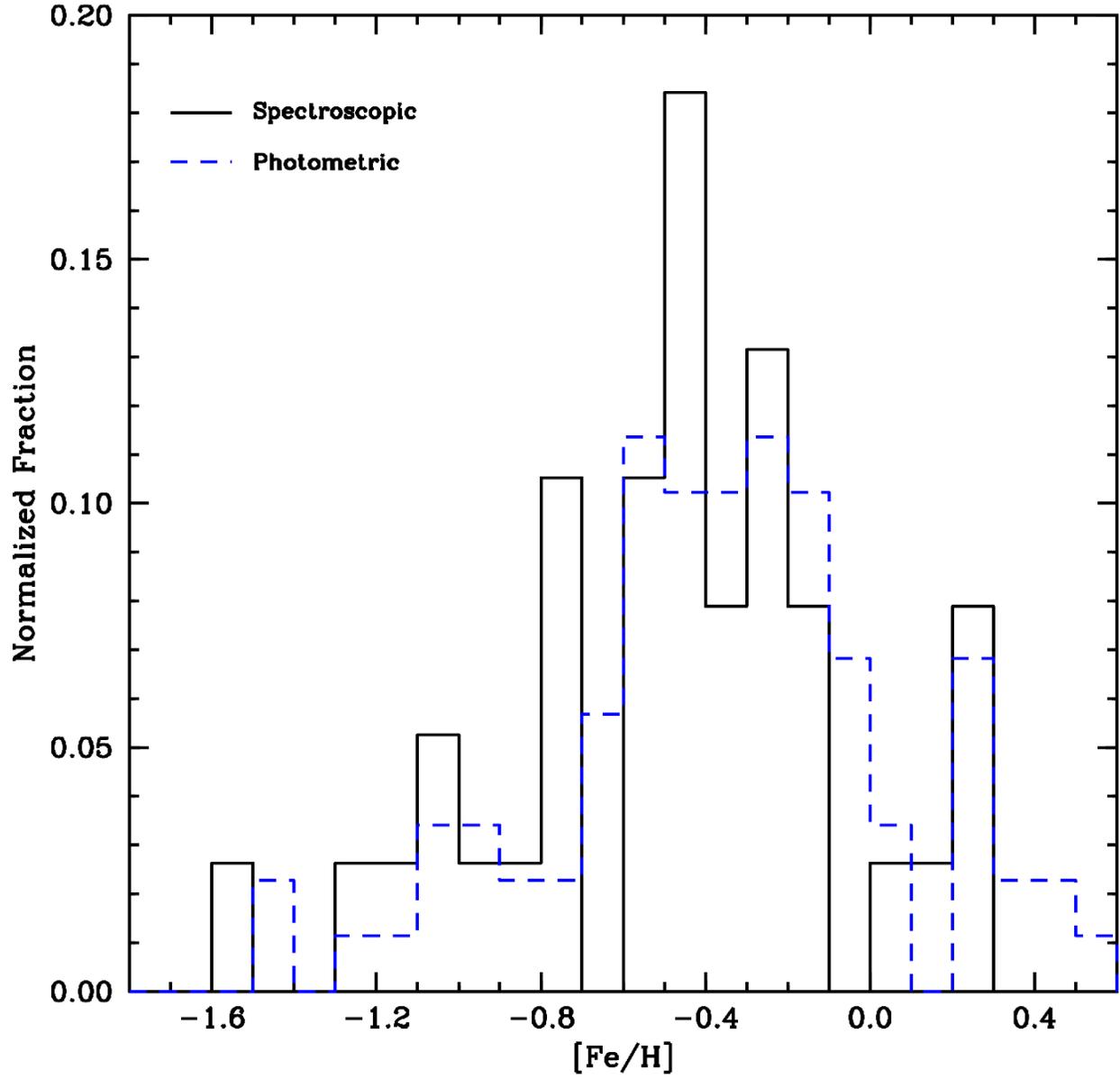}
\caption{Comparison of the metallicity distribution functions determined from
spectroscopy (solid black line) and photometry (dashed blue line) in 0.1 dex
bins.  The spectroscopic and photometric median [Fe/H] values are --0.41 and
--0.34, respectively.}
\label{f7}
\end{figure}

As mentioned in $\S$2 and evident in Figure \ref{f1} there is likely an 
observational bias in our derived metallicity distribution functions.  Although
the distribution functions are quite similar between the two target fields
(see Figure \ref{f6}), neither case samples the full color range of the bulge
RGB.  However, the combined RGB sample shown in Figure \ref{f6} exhibits
a full [Fe/H] range of $\sim$--1.5 to +0.3, which is nearly identical to 
previous bulge observations (e.g., Rich et al. 1988; Zoccali et al. 2008).  
Furthermore, simply dividing the RGB samples in half by 
(J--K$_{\rm s}$)$_{\rm o}$ color reveals that the average and median [Fe/H] 
differences are $<$0.10 dex between the redder and bluer stars.  Note that the 
same result is found if instead one divides the samples by fitting a line 
through the observed RGB stars on the color--magnitude diagram and compares
the average and median [Fe/H] ratios for stars lying above (brighter and bluer)
and below (dimmer and redder) the best--fit line.  We also show
in Figure \ref{f6} that the derived metallicity distribution function is
qualitatively in agreement with the simple, one--zone closed box enrichment 
model (e.g., Mould 1984; Rich 1990), which has been shown to be true in 
other bulge fields as well (e.g., Zoccali et al. 2003; 2008).  While these
observations suggest that our sample is not seriously biased toward the 
metal--poor end, it is likely that the most metal--rich stars are slightly 
underrepresented here.  In Figure \ref{f7} we directly compare
the metallicity distribution functions derived from spectroscopy (limited
color range) and photometry (full color range) and find that both 
distributions agree reasonably well in shape, median metallicity 
([Fe/H]$_{\rm spec.}$=--0.43; [Fe/H]$_{\rm phot.}$=--0.34), metallicity 
dispersion ($\sigma$$_{\rm spec.}$=0.42; $\sigma$$_{\rm phot.}$=0.49), and 
metallicity range (--1.5$\la$[Fe/H]$\la$+0.5).  Note also that both the 
photometric and spectroscopic metallicity distributions find a paucity of 
stars with [Fe/H]$<$--1.5, in agreement with the Zoccali et al. (2008) survey.

\begin{figure}
\epsscale{1.00}
\plotone{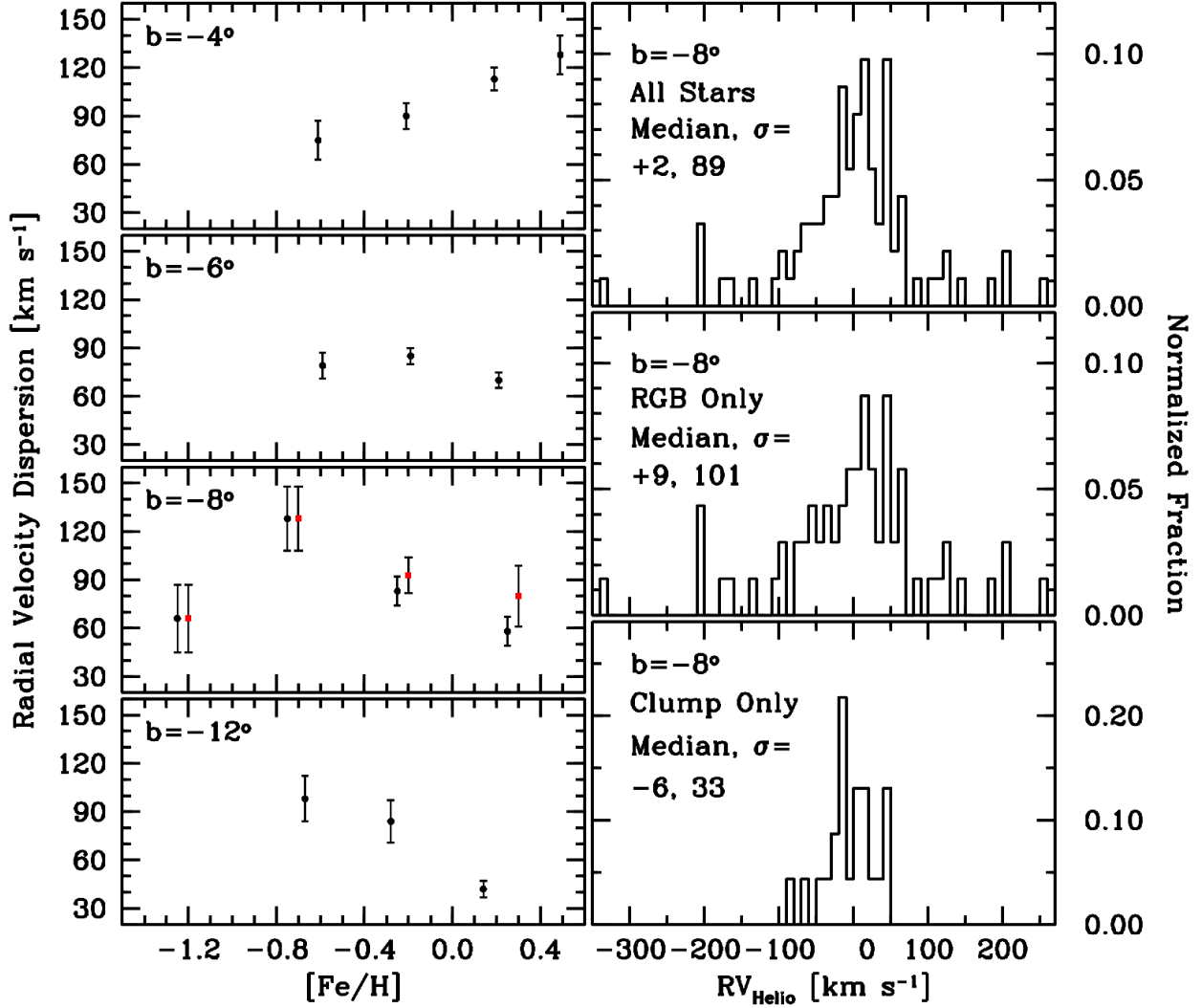}
\caption{The left panels show the measured radial velocity dispersion as a 
function of metallicity along the bulge minor axis at b=--4$\degr$, --6$\degr$, 
--8$\degr$, and --12$\degr$.  The b=--8$\degr$ data are from this study (0.5 
dex [Fe/H] bins), and the other fields were taken from Babusiaux et al. (2010;
0.4 dex bins).  In the b=--8$\degr$ panel the filled black circles represent 
the radial velocity dispersion when the full sample is taken into account, and 
the filled red squares represent the radial velocity dispersion when only the 
RGB stars are used.  The histograms in the right panels illustrate the radial
velocity distribution at b=--8$\degr$ with the full sample (top panel), the 
RGB stars only (middle panel), and the foreground red clump stars only (bottom 
panel) in 10 km s$^{\rm -1}$ bins.}
\label{f8}
\end{figure}

In Figure \ref{f8} we plot both the raw radial velocity distributions at
b=--8$\degr$ and the radial velocity dispersions binned by [Fe/H] in 0.5 dex
increments.  We also include the metallicity binned velocity dispersion data
from Babusiaux et al. (2010), which provide data at b=--4$\degr$, --6$\degr$, 
and --12$\degr$.  The median radial velocities and velocity dispersions 
measured here are in reasonable agreement with the \emph{BRAVA} M giant 
analysis of Plaut's field (Howard et al. 2009).  If all stars in our sample are
included then the median radial velocity is RV=+2 km s$^{\rm -1}$ ($\sigma$=89 
km s$^{\rm -1}$), which is similar to the \emph{BRAVA} results of 
$\langle$V$_{\rm GC}$$\rangle$$\approx$--20 km s$^{\rm -1}$ 
($\sigma$$\approx$90 km s$^{\rm -1}$).  However, if only the bulge RGB stars
are included then both the median radial velocity and dispersion increase
to RV=+9 km s$^{\rm -1}$ ($\sigma$=101 km s$^{\rm -1}$).  Fortunately, the
choice of whether to include the red clump stars, which exhibit a much lower
velocity dispersion ($\sigma$=33 km s$^{\rm -1}$; see Figure \ref{f8})
has little effect on the general shape of the Plaut field velocity dispersion
profile as a function of [Fe/H].  

In either case there is a clear decrease in velocity dispersion with increasing
metallicity.  This is similar to what Babusiaux et al. (2010) found at 
b=--12$\degr$, but appears quite different from the b=--6$\degr$ field, which 
exhibits no correlation with metallicity, and especially the b=--4$\degr$ field
where the most metal--rich stars appear to have the highest velocity 
dispersion.  Furthermore, the velocity dispersions found here and in the 
\emph{BRAVA} survey are at least 20--40 km s$^{\rm -1}$ larger than those 
predicted by the Zhao (1996) and Fux (1999) models, but the difference is 
$<$20 km s$^{\rm -1}$ compared with the recent Shen et al. (2010) model.  We 
caution the reader that the most metal--poor and metal--rich bins in our 
dataset contain $<$10 stars each, and additional observations may alter the 
velocity dispersion profile.  However, if the metallicity dependent velocity 
dispersion trends seen in the b=--8$\degr$ and --12$\degr$ fields are robust 
then we might expect the inclusion of more metal--rich stars to lower the 
overall observed velocity dispersion to be in better agreement with model 
predictions.

In addition to the 69 RGB stars we also serendipitously observed 23 stars 
that appear to be mostly well separated in color from the bulge giant branch
(see Figure \ref{f1}).  Similar stars along this vertical blue sequence in 
other bulge fields have been tentatively identified as disk red clump stars 
(e.g., Zoccali et al. 2003; Vieira et al. 2007; Rangwala et al. 2009), and the 
Besancon model\footnote{The Besancon model can be accessed at: 
http://model.obs-besancon.fr/.} (Robin et al. 2003; 2004) indicates that these 
stars should be mostly intermediate age ($\sim$2--7 Gyr) thin disk stars with 
a median [Fe/H]$\sim$+0.1, which is comparable to our derived [Fe/H]=+0.05.
In addition to being separated in color, these stars appear to have three 
interesting characteristics: (1) a similar median radial velocity to the RGB 
population but a velocity dispersion that is $\sim$3 times smaller, (2) a 
significantly higher median metallicity than the bulge and thick disk giants, 
and (3) noticeably enhanced [$\alpha$/Fe] ratios (see $\S$6.2).  

It is possible that the smaller velocity dispersion of these red clump stars 
may be due to the small sample size, and a two sided Kolmogorov--Smirnov (KS) 
test (Press et al. 1992) between the RGB and clump populations is unable to 
strongly reject the null hypothesis that these two groups are drawn from the 
same parent distribution.  To further test this we ran a simple bootstrap 
analysis with 10$^{\rm 6}$ trials that randomly selected 23 velocities from 
the RGB sample and measured the radial velocity dispersion.  We found that 
there is approximately a 0.004$\%$ chance that 23 RGB stars chosen at random 
would produce a velocity dispersion less than or equal to the red clump 
dispersion (33 km s$^{\rm -1}$).  However, this probability increases to 
$\sim$6$\%$ at the 2$\sigma$ level (66 km s$^{\rm -1}$).  

The metallicity distribution function appears somewhat less ambiguous.  A two 
sided KS test rejects the null hypothesis that the RGB and clump [Fe/H] 
distributions were drawn from the same parent population at the 99$\%$ level.
However, the same test does not rule out that the clump and thin disk stars
may share similar RV and [Fe/H] distributions (see also $\S$6.3).  Future 
observations with large sample sizes will be needed to fully investigate the 
true nature of these clump stars.

\subsection{Alpha Element Enhancement}

The consistent overproduction of $\alpha$ elements relative to the solar 
$\alpha$/Fe ratio in stellar populations is generally regarded as an indicator
of rapid ($\la$2x10$^{\rm 7}$ years) chemical enrichment associated with 
Type II SNe (e.g., Tinsley 1979; Matteucci \& Brocato 1990).  Furthermore,
it is believed that the onset of Type Ia SNe, which occurs on timescales 
$>$5$\times$10$^{\rm 8}$ years, drives the [$\alpha$/Fe] ratio downward 
because of the copious production of Fe--peak elements (e.g., Yoshii et al. 
1996; Nomoto et al. 1997).  While the downturn in the [$\alpha$/Fe] trend 
occurs near [Fe/H]$\approx$--1 for the thin disk (e.g., Bensby et al. 2003; 
Reddy et al. 2003; Brewer \& Carney 2006), the thick disk and bulge appear to 
retain $\alpha$--enhanced stars up to [Fe/H]$\approx$--0.3 (e.g., Fulbright
et al. 2007; Alves--Brito 2010).  This suggests that the thick disk and bulge 
experienced much more rapid and efficient star formation, but may also be an 
indication of other differences related to parameters such as the initial mass 
function, gas inflow/outflow, and the binary fraction.

Although past studies have shown that bulge stars generally maintain a high 
[$\alpha$/Fe] ratio, relative to the thin disk trend, at low Galactic latitudes
(McWilliam \& Rich 1994; Rich \& Origlia 2005; Cunha \& Smith 2006; Fulbright 
et al. 2007; Lecureur et al. 2007; Rich et al. 2007; Alves--Brito et al. 2010; 
Bensby et al. 2010), we provide here the first detailed analysis 
of stars approximately 1 kpc from the Galactic center.  In Figure \ref{f9} we 
plot our derived [Si/Fe], [Ca/Fe], and [$\alpha$/Fe]\footnote{Note that the 
[$\alpha$/Fe] value is the average of the [Si/Fe] and [Ca/Fe] ratios.} ratios 
as a function of [Fe/H], and compare our results to the thin and thick disk 
trends as well as Baade's window.  We find 
that the [$\alpha$/Fe] versus [Fe/H] trends in the Plaut field are nearly 
indistinguishable from those in Baade's window.  In other words, we find that 
no strong gradient in [$\alpha$/Fe] exists between b=--4$\degr$ and 
--8$\degr$, despite the clear presence of a metallicity gradient.  This 
suggests that the bulge chemical enrichment process acted rapidly and with 
surprising uniformity over a very large volume.  

Recent analyses have argued that the thick disk and bulge may share 
similar [$\alpha$/Fe] enhancements despite exhibiting clear differences in 
their metallicity distribution functions (Mel{\'e}ndez et al. 2008; 
Alves--Brito et al. 2010; Bensby et al. 2010; Ryde et al. 2010).  While we 
find that the [Si/Fe], [Ca/Fe], and [$\alpha$/Fe] ratios are generally
enhanced by $\sim$0.2 dex in bulge giants compared to thick disk giants (and 
dwarfs), this enhancement is not extraordinarily different than the combined 
measurement uncertainty and choice of abundance normalization scale mentioned 
in $\S$4.  Interestingly, both the thick disk and bulge appear to show similar 
declines in the [$\alpha$/Fe] ratio at [Fe/H]$\approx$--0.3.  However, it 
remains to be seen whether these two populations are truly distinct in their 
$\alpha$ element composition.  Fortunately, the difference in the [$\alpha$/Fe]
ratios between bulge and thin disk stars is much clearer, especially near the 
median metallicity of the Plaut field.  However, the two populations may appear
to merge with [$\alpha$/Fe]$\approx$0 at [Fe/H]$\ga$0.  More Plaut field 
observations at [Fe/H]$>$0 are needed to confirm the exact nature of the 
[$\alpha$/Fe] trend at the highest bulge metallicities.

Finally, in addition to the red clump stars shown in Figure \ref{f1} exhibiting
a high average metallicity, relatively small metallicity dispersion, and a 
low velocity dispersion, these stars appear to be just as enhanced in $\alpha$
elements as the bulge RGB stars.  We attribute the larger scatter in [Si/Fe],
[Ca/Fe], and [$\alpha$/Fe] to be due mostly to surface gravity uncertainties.
Nevertheless, these are likely foreground stars located $\sim$2--4 kpc away,
and their projected location at b=--8$\degr$ places them $\sim$300--600 pc 
below the Galactic plane.  However, the typical metallicities of these stars 
are much higher than most of the thick disk, and the [$\alpha$/Fe] 
enhancement suggests they do not share a similar composition with the more 
metal--rich thick and thin disk stars (see Figure \ref{f9}).  There is also 
the possibility that these stars, especially given their interesting kinematic 
and chemical composition properties, belong to a stellar stream, but 
unfortunately our current dataset is unable to test this scenario without 
additional observations.

\subsection{Possible RGB and Red Clump Contamination}

Since the bulge RGB stars analyzed here lie near the center of the Galaxy, 
roughly 8 kpc from the Sun, and 1 kpc below the Galactic plane, any sample that 
is chosen purely from a color--magnitude diagram will likely contain at least 
some Galactic thin disk, thick disk, and halo interlopers.  We first examined 
the likelihood of contamination by analyzing the expected number counts from
each population in a theoretical color--magnitude diagram, calculated from the
Besancon model, spanning the color and luminosity range of our targets.  The 
Besancon model returned the following population breakdown for the RGB sample:
3$\%$ young ($<$1 Gyr) thin disk, 33$\%$ old ($>$1 Gyr) thin disk, 20$\%$ thick
disk, 1$\%$ halo, and 43$\%$ bulge.  We also ran the simulation for the 
red clump stars and obtained the following distribution: 1$\%$ young thin disk,
62$\%$ old thin disk, 13$\%$ thick disk, 1$\%$ halo, and 23$\%$ bulge.  While
the halo can be effectively ruled out as a major contaminant, based on both
the [Fe/H] distribution and Besancon model, the thin and thick disk populations
require further analysis.

\begin{figure}
\epsscale{1.00}
\plotone{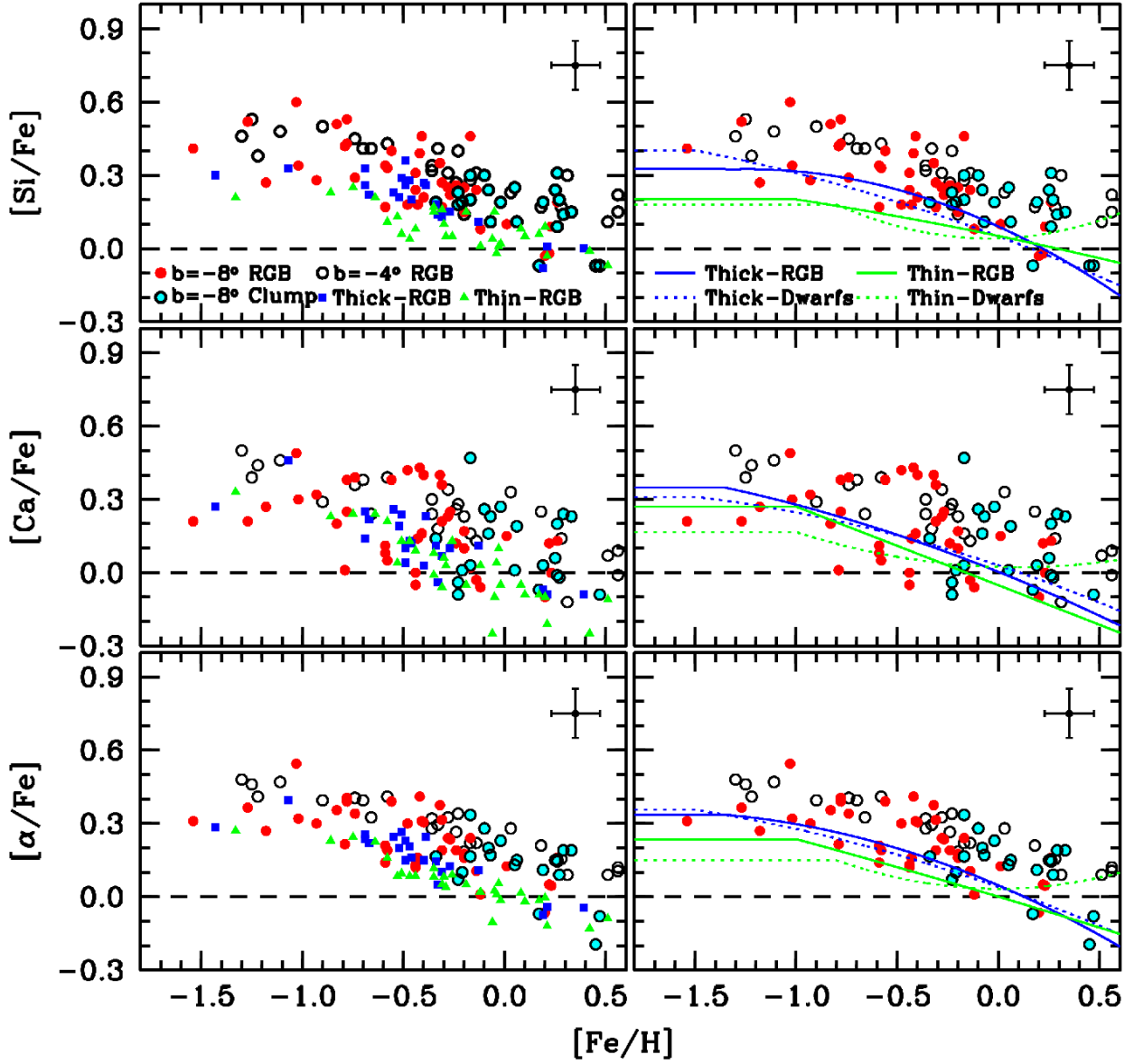}
\caption{The distribution of [Si/Fe], [Ca/Fe], and 
[$\alpha$/Fe] abundances as a function of [Fe/H].  The b=--4$\degr$ data
are from Fulbright et al. (2007; AODFNEW only), and the thick/thin disk giant 
data are from Alves--Brito et al. (2010; Kurucz model atmospheres).  The right 
panels show least squares fits to the literature thin and thick disk data for 
both giants (solid lines) and dwarfs (dotted lines).  The dwarf data were 
compiled from Fulbright et al. 2000, Bensby et al. (2003; 2005), and Reddy et 
al. (2003; 2006).}
\label{f9}
\end{figure}

\begin{figure}
\epsscale{1.00}
\plotone{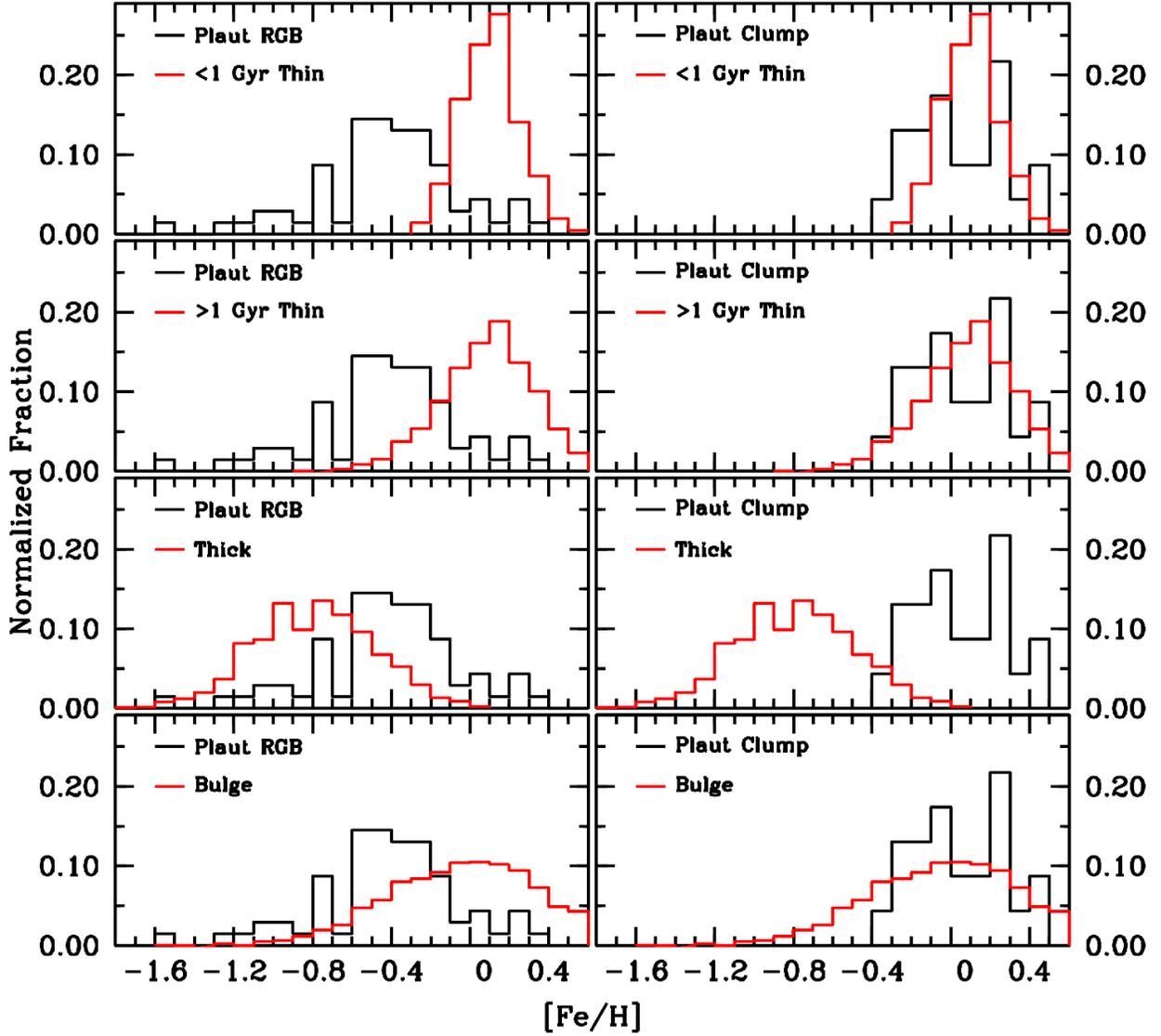}
\caption{Histograms of our observed [Fe/H] distributions (solid black lines)
versus those predicted for the young ($<$1 Gyr) thin disk, old ($>$1 Gyr) thin
disk, thick disk, and bulge by the Besancon model (solid red lines).  The 
left panels compare our RGB sample to the model and the right panels compare
our red clump sample to the model.}
\label{f10}
\end{figure}

\begin{figure}
\epsscale{1.00}
\plotone{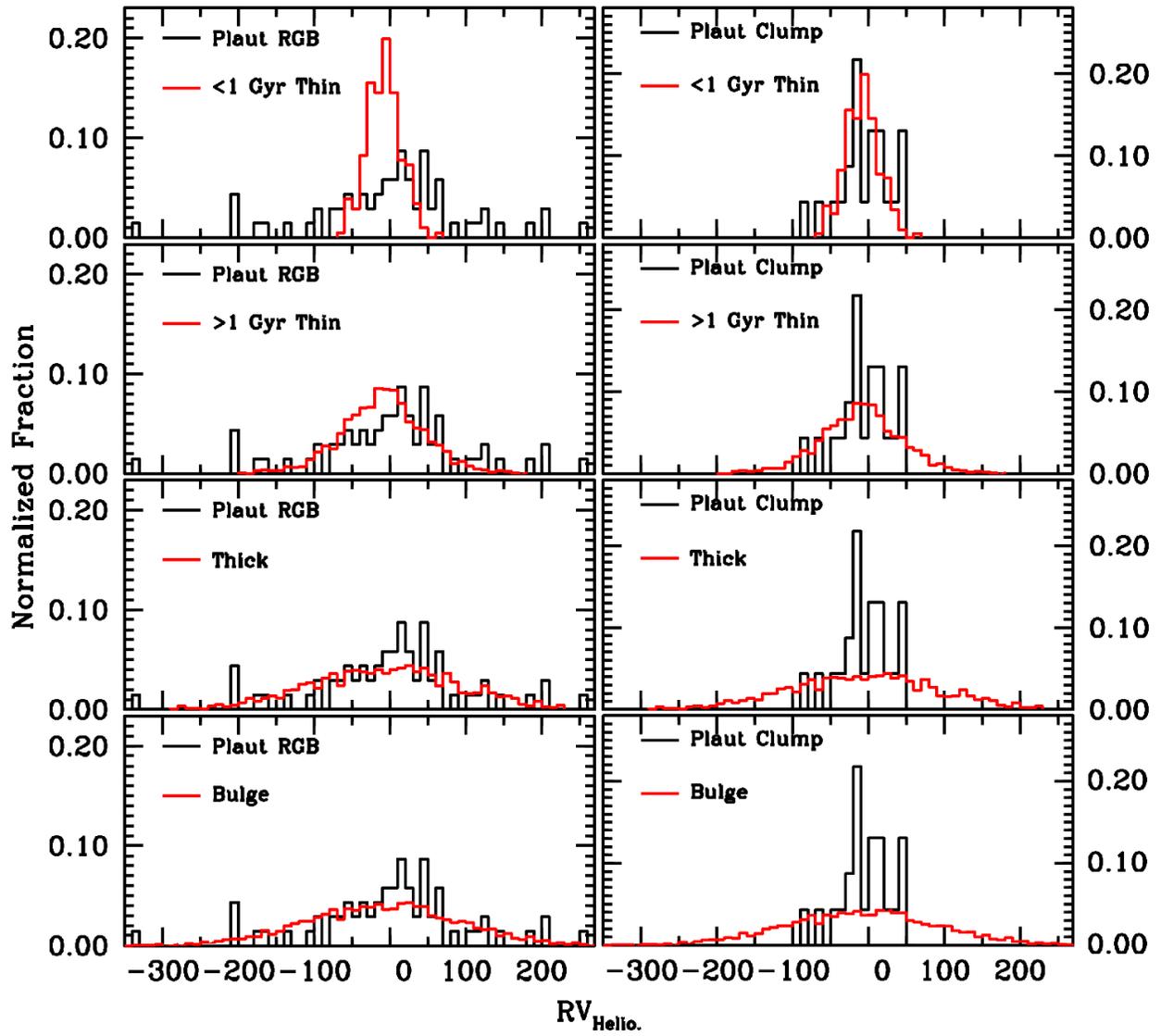}
\caption{Similar to Figure \ref{f10} but showing the radial velocity 
distributions.}
\label{f11}
\end{figure}

Figures \ref{f10}--\ref{f11} show our derived [Fe/H] and RV distributions 
compared with those predicted by the Besancon model for the young thin disk, 
old thin disk, thick disk, and bulge.  For the RGB sample the thin disk 
can be mostly ruled out as a major contaminant because the velocity and 
metallicity dispersions are too small, the average metallicity is too high, and
the majority of the Plaut stars have [$\alpha$/Fe] ratios that are at least a 
factor of two above those observed in the thin disk (see Figure \ref{f9}).  
On the other hand, the thick disk contamination is more difficult to rule out.
The thick disk appears to show a similar velocity distribution, but the median
[Fe/H] ratio of the Plaut field stars is $\sim$0.4 dex higher than in the 
thick disk.  Additionally, the Plaut stars extend to much higher metallicities
than the thick disk distribution.  However, thick disk contamination at about 
the 10--20$\%$ level for the more metal--poor Plaut RGB stars seems reasonable,
especially given that several ($\sim$5--10) of the Plaut RGB stars with 
[Fe/H]$<$--0.5 exhibit [$\alpha$/Fe] ratios that are in--line with literature
thick disk giants (see Figure \ref{f9}).  Interestingly, the theoretical bulge 
velocity distribution provided by the Besancon model fits our observed data
rather well, but the [Fe/H] distribution is a poor fit.  The model clearly
overestimates the average bulge metallicity at b=--8$\degr$, and instead 
predicts a distribution more similar to that seen at b=--4$\degr$.

For the foreground red clump sample the Besancon model effectively rules out 
the thick disk and bulge, based on both the velocity and [Fe/H] distributions. 
Although the average [Fe/H] value for the model bulge is similar to the red 
clump sample, we have already shown that the model is likely a poor fit to the 
true bulge metallicity distribution at b=--8$\degr$.  The thin disk provides a 
decent fit for both the [Fe/H] and velocity distributions, but the 
[$\alpha$/Fe] enhancements observed in the red clump stars are clearly above
those observed in the thin disk (see Figure \ref{f9}).  This reinforces the 
need for more observations of these objects.

\section{SUMMARY}

We have determined Fe, Si, and Ca abundances for 61 stars and Fe abundances for
an additional 31 stars in Plaut's low extinction window using high resolution 
(R$\approx$25,000) spectra obtained with the Hydra multifiber spectrograph at 
CTIO.  Additionally, we have derived a metallicity distribution function from 
2MASS infrared and optical photometry obtained with the Swope 40 inch telescope
at Las Campanas in a nearby bulge field that is in reasonable agreement with 
the spectroscopic results.  We find that the median [Fe/H] ratio has declined 
by $\sim$0.4 dex from Baade's Window at b=--4$\degr$, and that the 
b=--8$\degr$ field is part of a smoothly declining metallicity gradient.  
Despite a lower overall metallicity in the b=--8$\degr$ field, the full range 
of derived [Fe/H] abundances is nearly identical to previous studies at 
b=--4$\degr$, --6$\degr$, and --12$\degr$, with --1.5$<$[Fe/H]$<$+0.3.  
Furthermore, the metallicity distribution at b=--8$\degr$ is consistent with a 
simple model with a declining yield, due to the removal of metals by winds.  
The radial velocity distribution and dispersion of bulge RGB stars in the Plaut 
field are also in agreement with results from the \emph{BRAVA} survey, and we 
find that the velocity dispersion decreases as a function of increasing 
metallicity, similar to what was found previously at b=--12$\degr$.

The $\alpha$ element enhancements are consistent with a rapid enrichment 
process involving massive star SNe.  While the bulge RGB stars are enhanced
above the level of the thin and thick disk, the difference in [$\alpha$/Fe] 
between the bulge and thick disk is smaller and may be affected by systematic
offsets.  However, we find that the enhancements of Si and Ca are nearly
indistinguishable from those in Baade's Window, which suggests the lack of an
[$\alpha$/Fe] gradient along the bulge minor axis despite the presence of a 
metallicity gradient.  

A subset of our sample are candidate red clump stars that lie closer to the 
Sun, distributed along the line--of--sight toward the bulge.  While our data are
insufficient to assign a population membership to these stars, they exhibit
some very interesting properties that warrant further investigation: (1) the 
stars are significantly more metal--rich (median [Fe/H]$\sim$+0.05) than the 
bulge RGB sample and exhibit a rather small [Fe/H] dispersion, (2) the mean
radial velocity is similar to the bulge RGB stars but the velocity dispersion
is about a factor of three smaller, and (3) these stars also have enhanced
[$\alpha$/Fe] ratios like the bulge giants.  While similar stars in bulge 
color--magnitude diagrams have been designated as intermediate age disk red 
clump stars in the past, the data suggest that their chemical composition and 
kinematic properties are inconsistent with belonging to either the local thin 
or thick disk populations.

\acknowledgements

We would like to thank the anonymous referee for a careful reading and 
insightful comments that led to improvement of the manuscript.
We would also like to thank Mark Morris, Caty Pilachowski, and Andrea Kunder 
for helpful discussions, and David Reitzel for obtaining a portion of these 
observations.  This publication makes use of data products from the Two Micron 
All Sky Survey, which is a joint project of the University of Massachusetts and
the Infrared Processing and Analysis Center/California Institute of Technology,
funded by the National Aeronautics and Space Administration and the National 
Science Foundation.  This research has made use of NASA's Astrophysics Data 
System Bibliographic Services.  This research has made use of the NASA/IPAC 
Extragalactic Database (NED) which is operated by the Jet Propulsion 
Laboratory, California Institute of Technology, under contract with the 
National Aeronautics and Space Administration.  Support of the College of Arts 
and Sciences at Indiana University Bloomington for CIJ is gratefully 
acknowledged.  This material is based upon work supported by the 
National Science Foundation under award No. AST--1003201 to CIJ and award No. 
AST--0709479 to RMR.

\clearpage

\clearpage

\clearpage

\clearpage

\clearpage

\clearpage

\clearpage

\clearpage

\clearpage

\clearpage

\clearpage

\clearpage
\setlength{\hoffset}{-0.50in}

\tablenum{1}
\tablecolumns{12}
\tablewidth{0pt}

\begin{deluxetable}{cccccccccccc}
\tabletypesize{\scriptsize}
\tablecaption{Program Star Parameters and Results}
\tablehead{
\colhead{Star}	&
\colhead{V}	&
\colhead{J}      &
\colhead{K$_{\rm s}$}      &
\colhead{T$_{\rm eff}$}      &
\colhead{log g}      &
\colhead{[Fe/H]}      &
\colhead{V$_{\rm t}$}      &
\colhead{[Si/Fe]}      &
\colhead{[Ca/Fe]}      &
\colhead{RV$_{\rm Helio.}$}      &
\colhead{RGB/Field\tablenotemark{a}}      \\
\colhead{2MASS}  &
\colhead{}      &
\colhead{}      &
\colhead{}      &
\colhead{(K)}      &
\colhead{(cgs)}      &
\colhead{}      &
\colhead{(km s$^{\rm -1}$)}      &
\colhead{}      &
\colhead{}      &
\colhead{(km s$^{\rm -1}$)}      &
\colhead{}
}
\startdata
18174532$-$3353235	&	13.957	&	11.686	&	10.955	&	4540	&	1.45	&	$-$1.54	&	1.65	&	$+$0.41	&	$+$0.21	&	$-$16	&	RGB/1	\\
18182918$-$3341405	&	13.813	&	10.965	&	10.046	&	4125	&	0.95	&	$-$1.27	&	1.60	&	$+$0.52	&	$+$0.21	&	$+$51	&	RGB/1	\\
18175567$-$3343063	&	14.123	&	11.645	&	10.852	&	4425	&	1.30	&	$-$1.18	&	1.15	&	$+$0.27	&	$+$0.27	&	$-$107	&	RGB/1	\\
18181521$-$3352294	&	13.533	&	10.951	&	10.046	&	4215	&	1.00	&	$-$1.03	&	1.55	&	$+$0.60	&	$+$0.49	&	$+$30	&	RGB/1	\\
18182256$-$3401248	&	13.265	&	10.867	&	10.104	&	4465	&	1.10	&	$-$1.02	&	1.15	&	$+$0.34	&	$+$0.30	&	$+$49	&	RGB/1	\\
18174351$-$3401412	&	13.879	&	11.525	&	10.756	&	4450	&	1.35	&	$-$0.93	&	1.25	&	$+$0.28	&	$+$0.32	&	$+$2	&	RGB/1	\\
18182675$-$3248295	&	13.870	&	11.442	&	10.593	&	4435	&	1.25	&	$-$0.93	&	1.95	&	\nodata	&	\nodata	&	$-$36	&	RGB/2	\\
18172965$-$3402573	&	14.403	&	11.755	&	10.863	&	4155	&	1.35	&	$-$0.83	&	1.80	&	$+$0.51	&	$+$0.20	&	$+$23	&	RGB/1	\\
18183521$-$3344124	&	14.451	&	11.687	&	10.785	&	4150	&	1.25	&	$-$0.79	&	1.85	&	$+$0.42	&	$+$0.01	&	$-$25	&	RGB/1	\\
18181435$-$3350275	&	13.786	&	11.356	&	10.497	&	4340	&	1.20	&	$-$0.78	&	1.25	&	$+$0.43	&	$+$0.38	&	$+$2	&	RGB/1	\\
18183876$-$3403092	&	14.120	&	11.255	&	10.205	&	4000	&	1.05	&	$-$0.78	&	1.80	&	$+$0.53	&	$+$0.25	&	$+$12	&	RGB/1	\\
18175670$-$3246550	&	14.290	&	11.687	&	10.742	&	4290	&	1.30	&	$-$0.76	&	1.70	&	\nodata	&	\nodata	&	$+$66	&	RGB/2	\\
18174304$-$3357006	&	13.662	&	10.940	&	9.980	&	4090	&	1.00	&	$-$0.74	&	1.45	&	$+$0.29	&	$+$0.39	&	$+$209	&	RGB/1	\\
18182636$-$3253267	&	13.790	&	11.361	&	10.538	&	4445	&	1.25	&	$-$0.71	&	1.65	&	\nodata	&	\nodata	&	$-$331	&	RGB/2	\\
18173757$-$3256075	&	14.380	&	11.992	&	11.136	&	4560	&	1.50	&	$-$0.60	&	0.95	&	\nodata	&	\nodata	&	$+$208	&	RGB/2	\\
18180831$-$3405309	&	13.601	&	10.859	&	9.906	&	4100	&	1.00	&	$-$0.59	&	1.95	&	$+$0.34	&	$+$0.08	&	$-$95	&	RGB/1	\\
18180550$-$3407117	&	13.963	&	11.423	&	10.629	&	4335	&	1.25	&	$-$0.59	&	1.45	&	$+$0.17	&	$+$0.11	&	$-$96	&	RGB/1	\\
18185079$-$3259346	&	14.010	&	11.179	&	10.141	&	4055	&	1.05	&	$-$0.59	&	1.90	&	\nodata	&	\nodata	&	$+$10	&	RGB/2	\\
18174941$-$3353025	&	14.329	&	11.822	&	10.968	&	4275	&	1.40	&	$-$0.58	&	1.95	&	$+$0.33	&	$+$0.05	&	$-$74	&	RGB/1	\\
18184795$-$3257096	&	14.130	&	11.671	&	10.765	&	4360	&	1.35	&	$-$0.58	&	1.75	&	\nodata	&	\nodata	&	$+$45	&	RGB/2	\\
18174742$-$3348098	&	14.710	&	11.984	&	11.082	&	4130	&	1.45	&	$-$0.56	&	1.45	&	$+$0.40	&	$+$0.38	&	$-$18	&	RGB/1	\\
18185907$-$3249511	&	14.360	&	11.745	&	10.828	&	4255	&	1.35	&	$-$0.55	&	1.85	&	\nodata	&	\nodata	&	$-$77	&	RGB/2	\\
18184583$-$3240045	&	14.180	&	11.911	&	11.046	&	4540	&	1.45	&	$-$0.53	&	1.05	&	\nodata	&	\nodata	&	$+$6	&	RGB/2	\\
18180303$-$3256322	&	14.350	&	11.639	&	10.621	&	4150	&	1.25	&	$-$0.52	&	2.05	&	\nodata	&	\nodata	&	$+$184	&	RGB/2	\\
18183679$-$3251454	&	14.170	&	11.246	&	10.234	&	4025	&	1.10	&	$-$0.51	&	2.00	&	\nodata	&	\nodata	&	$-$210	&	RGB/2	\\
18181929$-$3404128	&	13.729	&	10.825	&	9.852	&	4020	&	0.90	&	$-$0.48	&	1.30	&	$+$0.18	&	$+$0.42	&	$+$61	&	RGB/1	\\
18185744$-$3247526	&	14.010	&	11.428	&	10.510	&	4275	&	1.20	&	$-$0.46	&	1.85	&	\nodata	&	\nodata	&	$+$63	&	RGB/2	\\
18181512$-$3353545	&	14.463	&	11.895	&	11.067	&	4270	&	1.45	&	$-$0.44	&	1.65	&	$+$0.24	&	$+$0.00	&	$+$117	&	RGB/1	\\
18183802$-$3355441	&	13.983	&	11.446	&	10.639	&	4325	&	1.25	&	$-$0.44	&	1.75	&	$+$0.31	&	$-$0.05	&	$-$58	&	RGB/1	\\
18174303$-$3355118	&	14.265	&	11.708	&	10.813	&	4220	&	1.30	&	$-$0.43	&	1.60	&	$+$0.18	&	$+$0.14	&	$-$29	&	RGB/1	\\
18181659$-$3252450	&	14.440	&	12.046	&	11.199	&	4465	&	1.85	&	$-$0.43	&	2.15	&	\nodata	&	\nodata	&	$-$179	&	RGB/2	\\
18182470$-$3342166	&	14.069	&	11.284	&	10.353	&	4145	&	1.10	&	$-$0.42	&	1.60	&	$+$0.39	&	$+$0.43	&	$+$44	&	RGB/1	\\
18180285$-$3342004	&	14.570	&	11.754	&	10.835	&	4140	&	1.30	&	$-$0.41	&	1.45	&	$+$0.46	&	$+$0.16	&	$+$27	&	RGB/1	\\
18181783$-$3300021	&	14.200	&	11.743	&	10.890	&	4410	&	1.40	&	$-$0.41	&	1.15	&	\nodata	&	\nodata	&	$+$50	&	RGB/2	\\
18174935$-$3404217	&	13.777	&	11.340	&	10.523	&	4360	&	1.25	&	$-$0.40	&	1.70	&	$+$0.21	&	$+$0.40	&	$-$38	&	RGB/1	\\
18180218$-$3241380	&	14.410	&	12.243	&	11.397	&	4665	&	2.00	&	$-$0.39	&	1.95	&	\nodata	&	\nodata	&	$+$45	&	RGB/2	\\
18173180$-$3300124	&	13.960	&	11.183	&	10.181	&	4200	&	1.10	&	$-$0.38	&	1.70	&	\nodata	&	\nodata	&	$+$122	&	RGB/2	\\
18182720$-$3254318	&	14.280	&	11.699	&	10.812	&	4290	&	1.35	&	$-$0.38	&	1.30	&	\nodata	&	\nodata	&	$+$254	&	RGB/2	\\
18185164$-$3243594	&	14.280	&	12.030	&	11.219	&	4600	&	1.95	&	$-$0.35	&	1.60	&	\nodata	&	\nodata	&	$+$28	&	RGB/2	\\
18174929$-$3347192	&	14.046	&	12.006	&	11.362	&	4875	&	2.35	&	$-$0.34	&	1.40	&	$+$0.19	&	$+$0.14	&	$+$42	&	Clump/1	\\
18183604$-$3342349	&	13.431	&	10.573	&	9.649	&	4105	&	0.80	&	$-$0.32	&	1.90	&	$+$0.35	&	$+$0.40	&	$-$59	&	RGB/1	\\
18183644$-$3254249	&	14.390	&	11.684	&	10.712	&	4160	&	1.30	&	$-$0.32	&	1.55	&	\nodata	&	\nodata	&	$-$138	&	RGB/2	\\
18180562$-$3346548	&	14.293	&	11.638	&	10.782	&	4220	&	1.30	&	$-$0.31	&	1.75	&	$+$0.27	&	$+$0.36	&	$+$41	&	RGB/1	\\
18173554$-$3405009	&	13.674	&	11.528	&	10.772	&	4625	&	2.10	&	$-$0.31	&	1.60	&	$+$0.17	&	$+$0.21	&	$+$81	&	RGB/1	\\
18185946$-$3240274	&	13.960	&	11.344	&	10.322	&	4200	&	1.15	&	$-$0.31	&	1.50	&	\nodata	&	\nodata	&	$+$36	&	RGB/2	\\
18173180$-$3349197	&	14.328	&	11.464	&	10.494	&	4005	&	1.15	&	$-$0.28	&	1.25	&	$+$0.25	&	$+$0.23	&	$+$17	&	RGB/1	\\
18181924$-$3350222	&	14.360	&	11.761	&	10.887	&	4215	&	1.35	&	$-$0.27	&	1.55	&	$+$0.22	&	$+$0.25	&	$-$8	&	RGB/1	\\
18182065$-$3248452	&	14.200	&	11.567	&	10.636	&	4240	&	1.25	&	$-$0.27	&	2.10	&	\nodata	&	\nodata	&	$-$209	&	RGB/2	\\
18175593$-$3400000	&	14.053	&	11.380	&	10.479	&	4160	&	1.20	&	$-$0.24	&	1.40	&	$+$0.26	&	$+$0.12	&	$+$14	&	RGB/1	\\
18182089$-$3348425	&	14.284	&	12.129	&	11.428	&	4700	&	2.15	&	$-$0.23	&	1.45	&	$+$0.23	&	$-$0.09	&	$-$43	&	Clump/1	\\
18175546$-$3404103	&	12.960	&	11.012	&	10.351	&	4950	&	2.40	&	$-$0.23	&	1.40	&	$+$0.18	&	$-$0.04	&	$-$61	&	Clump/1	\\
18174688$-$3257530	&	14.260	&	11.981	&	11.170	&	4665	&	2.00	&	$-$0.22	&	1.80	&	\nodata	&	\nodata	&	$+$11	&	RGB/2	\\
18174798$-$3359361	&	13.733	&	11.721	&	11.038	&	4830	&	2.30	&	$-$0.21	&	1.75	&	$+$0.19	&	$+$0.01	&	$-$11	&	Clump/1	\\
18184297$-$3248086	&	14.300	&	11.603	&	10.622	&	4165	&	1.25	&	$-$0.21	&	1.35	&	\nodata	&	\nodata	&	$+$123	&	RGB/2	\\
18180979$-$3351416	&	13.107	&	10.679	&	9.832	&	4350	&	1.65	&	$-$0.20	&	1.70	&	$+$0.25	&	$+$0.10	&	$+$19	&	RGB/1	\\
18182430$-$3352453	&	13.900	&	11.163	&	10.265	&	4130	&	1.05	&	$-$0.20	&	1.40	&	$+$0.15	&	$+$0.17	&	$+$22	&	RGB/1	\\
18182494$-$3250309	&	14.260	&	12.117	&	11.241	&	4650	&	2.00	&	$-$0.20	&	1.35	&	\nodata	&	\nodata	&	$-$8	&	RGB/2	\\
18173251$-$3354539	&	14.271	&	12.033	&	11.304	&	4565	&	1.85	&	$-$0.17	&	2.10	&	$+$0.46	&	$+$0.02	&	$-$6	&	RGB/1	\\
18181710$-$3401088	&	14.395	&	12.342	&	11.747	&	4940	&	2.40	&	$-$0.17	&	1.45	&	$+$0.20	&	$+$0.47	&	$+$0	&	Clump/1	\\
18182553$-$3349465	&	14.281	&	12.094	&	11.383	&	4660	&	2.00	&	$-$0.17	&	1.75	&	$+$0.30	&	$+$0.03	&	$+$35	&	Clump/1	\\
18180991$-$3403206	&	13.839	&	11.327	&	10.517	&	4345	&	1.80	&	$-$0.14	&	1.75	&	$+$0.24	&	$-$0.03	&	$+$62	&	RGB/1	\\
18184496$-$3256146	&	14.160	&	11.597	&	10.674	&	4275	&	1.30	&	$-$0.14	&	2.05	&	\nodata	&	\nodata	&	$-$163	&	RGB/2	\\
18184867$-$3242133	&	13.930	&	11.445	&	10.542	&	4345	&	1.25	&	$-$0.13	&	1.90	&	\nodata	&	\nodata	&	$-$44	&	RGB/2	\\
18180301$-$3405313	&	13.896	&	11.011	&	10.073	&	4040	&	1.00	&	$-$0.12	&	2.05	&	$+$0.08	&	$-$0.06	&	$-$53	&	RGB/1	\\
18174900$-$3247128	&	14.450	&	11.523	&	10.468	&	4070	&	1.15	&	$-$0.11	&	2.30	&	\nodata	&	\nodata	&	$+$41	&	RGB/2	\\
18180502$-$3355071	&	13.877	&	11.742	&	11.049	&	4715	&	2.35	&	$-$0.10	&	1.60	&	$+$0.30	&	$+$0.26	&	$+$44	&	Clump/1	\\
18182740$-$3356447	&	14.387	&	12.172	&	11.505	&	4690	&	2.35	&	$-$0.09	&	1.55	&	\nodata	&	\nodata	&	$+$11	&	Clump/1	\\
18182457$-$3344533	&	14.158	&	11.823	&	11.055	&	4530	&	2.10	&	$-$0.08	&	1.90	&	$+$0.24	&	$+$0.15	&	$-$15	&	Clump/1	\\
18182612$-$3353431	&	14.319	&	12.283	&	11.625	&	4860	&	2.40	&	$-$0.07	&	1.65	&	$+$0.11	&	$+$0.23	&	$-$23	&	Clump/1	\\
18174386$-$3244555	&	13.850	&	11.113	&	10.114	&	4200	&	1.35	&	$-$0.07	&	1.90	&	\nodata	&	\nodata	&	$+$148	&	RGB/2	\\
18172979$-$3401118	&	13.507	&	11.507	&	10.846	&	4860	&	2.40	&	$-$0.02	&	1.40	&	$+$0.19	&	$+$0.27	&	$+$25	&	Clump/1	\\
18174571$-$3259300	&	14.320	&	12.053	&	11.288	&	4720	&	2.05	&	$+$0.00	&	1.35	&	\nodata	&	\nodata	&	$+$101	&	RGB/2	\\
18183930$-$3353425	&	14.466	&	11.833	&	10.936	&	4200	&	1.40	&	$+$0.01	&	1.75	&	$+$0.10	&	$+$0.15	&	$-$66	&	RGB/1	\\
18181293$-$3240588	&	14.070	&	11.752	&	10.869	&	4485	&	2.00	&	$+$0.01	&	1.30	&	\nodata	&	\nodata	&	$-$200	&	RGB/2	\\
18174891$-$3406031	&	13.505	&	11.457	&	10.774	&	4805	&	2.40	&	$+$0.05	&	1.70	&	$+$0.25	&	$+$0.01	&	$-$13	&	Clump/1	\\
18181322$-$3402227	&	13.350	&	11.186	&	10.507	&	4740	&	2.40	&	$+$0.06	&	1.50	&	$+$0.11	&	$+$0.19	&	$-$6	&	Clump/1	\\
18185947$-$3246054	&	14.310	&	11.425	&	10.462	&	4075	&	1.15	&	$+$0.07	&	1.50	&	\nodata	&	\nodata	&	$-$70	&	RGB/2	\\
18181033$-$3352390	&	13.787	&	11.793	&	11.146	&	4900	&	2.50	&	$+$0.17	&	1.30	&	$-$0.07	&	$-$0.07	&	$-$18	&	Clump/1	\\
18174000$-$3406266	&	13.254	&	10.998	&	10.279	&	4565	&	2.30	&	$+$0.19	&	1.85	&	$+$0.19	&	$+$0.03	&	$-$88	&	Clump/1	\\
18180012$-$3358096	&	13.848	&	11.066	&	10.144	&	4090	&	1.50	&	$+$0.20	&	1.95	&	$-$0.03	&	$-$0.10	&	$-$12	&	RGB/1	\\
18175652$-$3347050	&	13.855	&	11.183	&	10.293	&	4185	&	1.65	&	$+$0.22	&	1.40	&	$-$0.02	&	$+$0.12	&	$-$2	&	RGB/1	\\
18182472$-$3352044	&	13.858	&	11.043	&	10.117	&	4070	&	1.45	&	$+$0.23	&	1.80	&	$+$0.09	&	$+$0.00	&	$+$9	&	RGB/1	\\
18173706$-$3405569	&	14.581	&	12.126	&	11.409	&	4405	&	2.20	&	$+$0.25	&	2.10	&	$+$0.24	&	$+$0.06	&	$+$13	&	Clump/1	\\
18173994$-$3358331	&	14.774	&	11.962	&	11.079	&	4085	&	1.45	&	$+$0.26	&	1.65	&	$+$0.19	&	$+$0.13	&	$-$41	&	RGB/1	\\
18182073$-$3353250	&	14.509	&	12.185	&	11.425	&	4500	&	2.30	&	$+$0.26	&	1.55	&	$+$0.31	&	$-$0.01	&	$+$1	&	Clump/1	\\
18174478$-$3343290	&	14.047	&	11.885	&	11.224	&	4785	&	2.40	&	$+$0.26	&	1.85	&	$+$0.09	&	$+$0.20	&	$+$41	&	Clump/1	\\
18183369$-$3352038	&	13.758	&	11.523	&	10.770	&	4585	&	2.35	&	$+$0.27	&	1.55	&	$+$0.20	&	$-$0.02	&	$+$10	&	Clump/1	\\
18183098$-$3358070	&	13.511	&	11.197	&	10.484	&	4570	&	2.35	&	$+$0.29	&	1.65	&	$+$0.14	&	$+$0.23	&	$+$17	&	Clump/1	\\
18180049$-$3246462	&	14.250	&	11.821	&	10.956	&	4460	&	2.15	&	$+$0.32	&	1.45	&	\nodata	&	\nodata	&	$-$34	&	RGB/2	\\
18174067$-$3356000	&	12.582	&	10.319	&	9.577	&	4540	&	2.35	&	$+$0.33	&	1.70	&	$+$0.15	&	$+$0.23	&	$-$40	&	Clump/1	\\
18173118$-$3358318	&	14.225	&	11.685	&	10.928	&	4355	&	2.00	&	$+$0.45	&	1.85	&	$-$0.07	&	$-$0.32	&	$-$13	&	Clump/1	\\
18182052$-$3345251	&	13.800	&	11.447	&	10.669	&	4505	&	2.30	&	$+$0.47	&	1.75	&	$-$0.07	&	$-$0.09	&	$-$25	&	Clump/1	\\
\enddata

\tablenotetext{a}{``RGB": probable RGB member; ``Clump": probable red clump 
member; `1': Field 1 (l=--1$\degr$,b=--8.5$\degr$); `2': Field 2 
(l=0$\degr$,b=--8$\degr$)}

\end{deluxetable}

\clearpage

\tablenum{2}
\tablecolumns{7}
\tablewidth{0pt}

\begin{deluxetable}{ccccccc}
\tabletypesize{\scriptsize}
\tablecaption{Random and Systematic Uncertainties}
\tablehead{
\colhead{Star}	&
\colhead{$\sigma$$_{\rm Rand.}$}	&
\colhead{$\sigma$$_{\rm Sys.}$}      &
\colhead{$\sigma$$_{\rm Rand.}$}      &
\colhead{$\sigma$$_{\rm Sys.}$}      &
\colhead{$\sigma$$_{\rm Rand.}$}      &
\colhead{$\sigma$$_{\rm Sys.}$}      \\
\colhead{2MASS}    &
\colhead{log $\epsilon$(Fe)}	&
\colhead{log $\epsilon$(Fe)}    &
\colhead{log $\epsilon$(Si)}    &
\colhead{log $\epsilon$(Si)}    &
\colhead{log $\epsilon$(Ca)}    &
\colhead{log $\epsilon$(Ca)} 
}
\startdata
18174532$-$3353235	&	0.15	&	0.06	&	0.18	&	0.04	&	0.02	&	0.06	\\
18182918$-$3341405	&	0.14	&	0.06	&	0.00	&	0.07	&	0.11	&	0.07	\\
18175567$-$3343063	&	0.10	&	0.07	&	0.06	&	0.06	&	0.14	&	0.06	\\
18181521$-$3352294	&	0.11	&	0.06	&	0.06	&	0.09	&	0.10	&	0.08	\\
18182256$-$3401248	&	0.15	&	0.07	&	0.00	&	0.06	&	0.00	&	0.07	\\
18174351$-$3401412	&	0.16	&	0.07	&	0.00	&	0.08	&	0.11	&	0.07	\\
18182675$-$3248295	&	0.18	&	0.05	&	\nodata	&	\nodata	&	\nodata	&	\nodata	\\
18172965$-$3402573	&	0.12	&	0.08	&	0.06	&	0.11	&	0.12	&	0.08	\\
18183521$-$3344124	&	0.13	&	0.08	&	0.12	&	0.10	&	0.11	&	0.07	\\
18181435$-$3350275	&	0.14	&	0.06	&	0.07	&	0.09	&	0.14	&	0.08	\\
18183876$-$3403092	&	0.14	&	0.09	&	0.06	&	0.12	&	0.11	&	0.09	\\
18175670$-$3246550	&	0.21	&	0.07	&	\nodata	&	\nodata	&	\nodata	&	\nodata	\\
18174304$-$3357006	&	0.15	&	0.09	&	0.09	&	0.10	&	0.14	&	0.09	\\
18182636$-$3253267	&	0.16	&	0.06	&	\nodata	&	\nodata	&	\nodata	&	\nodata	\\
18173757$-$3256075	&	0.32	&	0.08	&	\nodata	&	\nodata	&	\nodata	&	\nodata	\\
18180831$-$3405309	&	0.15	&	0.07	&	0.05	&	0.09	&	0.09	&	0.08	\\
18180550$-$3407117	&	0.13	&	0.07	&	0.08	&	0.07	&	0.14	&	0.08	\\
18185079$-$3259346	&	0.24	&	0.08	&	\nodata	&	\nodata	&	\nodata	&	\nodata	\\
18174941$-$3353025	&	0.13	&	0.07	&	0.13	&	0.09	&	0.06	&	0.07	\\
18184795$-$3257096	&	0.24	&	0.06	&	\nodata	&	\nodata	&	\nodata	&	\nodata	\\
18174742$-$3348098	&	0.13	&	0.08	&	0.08	&	0.09	&	0.19	&	0.10	\\
18185907$-$3249511	&	0.08	&	0.07	&	\nodata	&	\nodata	&	\nodata	&	\nodata	\\
18184583$-$3240045	&	0.23	&	0.08	&	\nodata	&	\nodata	&	\nodata	&	\nodata	\\
18180303$-$3256322	&	0.32	&	0.07	&	\nodata	&	\nodata	&	\nodata	&	\nodata	\\
18183679$-$3251454	&	0.18	&	0.07	&	\nodata	&	\nodata	&	\nodata	&	\nodata	\\
18181929$-$3404128	&	0.21	&	0.10	&	0.00	&	0.10	&	0.04	&	0.10	\\
18185744$-$3247526	&	0.15	&	0.06	&	\nodata	&	\nodata	&	\nodata	&	\nodata	\\
18181512$-$3353545	&	0.10	&	0.07	&	0.07	&	0.08	&	0.13	&	0.08	\\
18183802$-$3355441	&	0.11	&	0.07	&	0.24	&	0.08	&	0.05	&	0.08	\\
18174303$-$3355118	&	0.16	&	0.08	&	0.10	&	0.10	&	0.14	&	0.08	\\
18181659$-$3252450	&	0.20	&	0.06	&	\nodata	&	\nodata	&	\nodata	&	\nodata	\\
18182470$-$3342166	&	0.17	&	0.09	&	0.06	&	0.11	&	0.02	&	0.10	\\
18180285$-$3342004	&	0.13	&	0.09	&	0.00	&	0.13	&	0.07	&	0.10	\\
18181783$-$3300021	&	0.17	&	0.08	&	\nodata	&	\nodata	&	\nodata	&	\nodata	\\
18174935$-$3404217	&	0.14	&	0.08	&	0.01	&	0.09	&	0.14	&	0.08	\\
18180218$-$3241380	&	0.17	&	0.05	&	\nodata	&	\nodata	&	\nodata	&	\nodata	\\
18173180$-$3300124	&	0.15	&	0.09	&	\nodata	&	\nodata	&	\nodata	&	\nodata	\\
18182720$-$3254318	&	0.27	&	0.09	&	\nodata	&	\nodata	&	\nodata	&	\nodata	\\
18185164$-$3243594	&	0.21	&	0.06	&	\nodata	&	\nodata	&	\nodata	&	\nodata	\\
18174929$-$3347192	&	0.19	&	0.06	&	0.05	&	0.07	&	0.10	&	0.07	\\
18183604$-$3342349	&	0.15	&	0.10	&	0.03	&	0.12	&	0.06	&	0.10	\\
18183644$-$3254249	&	0.17	&	0.10	&	\nodata	&	\nodata	&	\nodata	&	\nodata	\\
18180562$-$3346548	&	0.14	&	0.09	&	0.06	&	0.11	&	0.06	&	0.10	\\
18173554$-$3405009	&	0.17	&	0.06	&	0.05	&	0.08	&	0.00	&	0.07	\\
18185946$-$3240274	&	0.22	&	0.09	&	\nodata	&	\nodata	&	\nodata	&	\nodata	\\
18173180$-$3349197	&	0.15	&	0.12	&	0.16	&	0.13	&	0.12	&	0.10	\\
18181924$-$3350222	&	0.14	&	0.09	&	0.20	&	0.12	&	0.09	&	0.10	\\
18182065$-$3248452	&	0.17	&	0.08	&	\nodata	&	\nodata	&	\nodata	&	\nodata	\\
18175593$-$3400000	&	0.14	&	0.10	&	0.12	&	0.11	&	0.06	&	0.10	\\
18182089$-$3348425	&	0.16	&	0.06	&	0.07	&	0.09	&	0.15	&	0.07	\\
18175546$-$3404103	&	0.14	&	0.06	&	0.00	&	0.06	&	0.10	&	0.06	\\
18174688$-$3257530	&	0.17	&	0.06	&	\nodata	&	\nodata	&	\nodata	&	\nodata	\\
18174798$-$3359361	&	0.12	&	0.06	&	0.04	&	0.07	&	0.00	&	0.07	\\
18184297$-$3248086	&	0.29	&	0.10	&	\nodata	&	\nodata	&	\nodata	&	\nodata	\\
18180979$-$3351416	&	0.09	&	0.08	&	0.05	&	0.10	&	0.10	&	0.08	\\
18182430$-$3352453	&	0.15	&	0.10	&	0.07	&	0.11	&	0.09	&	0.10	\\
18182494$-$3250309	&	0.29	&	0.07	&	\nodata	&	\nodata	&	\nodata	&	\nodata	\\
18173251$-$3354539	&	0.10	&	0.06	&	0.09	&	0.10	&	0.12	&	0.07	\\
18181710$-$3401088	&	0.15	&	0.06	&	0.14	&	0.07	&	0.15	&	0.08	\\
18182553$-$3349465	&	0.21	&	0.06	&	0.04	&	0.08	&	0.01	&	0.07	\\
18180991$-$3403206	&	0.19	&	0.08	&	0.09	&	0.09	&	0.06	&	0.08	\\
18184496$-$3256146	&	0.14	&	0.08	&	\nodata	&	\nodata	&	\nodata	&	\nodata	\\
18184867$-$3242133	&	0.18	&	0.07	&	\nodata	&	\nodata	&	\nodata	&	\nodata	\\
18180301$-$3405313	&	0.13	&	0.09	&	0.08	&	0.11	&	0.02	&	0.09	\\
18174900$-$3247128	&	0.16	&	0.08	&	\nodata	&	\nodata	&	\nodata	&	\nodata	\\
18180502$-$3355071	&	0.12	&	0.07	&	0.10	&	0.09	&	0.09	&	0.07	\\
18182740$-$3356447	&	0.17	&	0.07	&	0.00	&	\nodata	&	0.00	&	\nodata	\\
18182457$-$3344533	&	0.16	&	0.06	&	0.05	&	0.08	&	0.15	&	0.08	\\
18182612$-$3353431	&	0.12	&	0.07	&	0.09	&	0.07	&	0.11	&	0.07	\\
18174386$-$3244555	&	0.21	&	0.08	&	\nodata	&	\nodata	&	\nodata	&	\nodata	\\
18172979$-$3401118	&	0.15	&	0.07	&	0.04	&	0.08	&	0.13	&	0.09	\\
18174571$-$3259300	&	0.19	&	0.07	&	\nodata	&	\nodata	&	\nodata	&	\nodata	\\
18183930$-$3353425	&	0.17	&	0.09	&	0.11	&	0.09	&	0.14	&	0.10	\\
18181293$-$3240588	&	0.10	&	0.08	&	\nodata	&	\nodata	&	\nodata	&	\nodata	\\
18174891$-$3406031	&	0.12	&	0.07	&	0.07	&	0.08	&	0.10	&	0.07	\\
18181322$-$3402227	&	0.11	&	0.08	&	0.11	&	0.08	&	0.00	&	0.08	\\
18185947$-$3246054	&	0.27	&	0.11	&	\nodata	&	\nodata	&	\nodata	&	\nodata	\\
18181033$-$3352390	&	0.15	&	0.07	&	0.07	&	0.08	&	0.12	&	0.08	\\
18174000$-$3406266	&	0.12	&	0.09	&	0.03	&	0.11	&	0.05	&	0.09	\\
18180012$-$3358096	&	0.13	&	0.12	&	0.15	&	0.13	&	0.00	&	0.09	\\
18175652$-$3347050	&	0.12	&	0.11	&	0.10	&	0.12	&	0.04	&	0.11	\\
18182472$-$3352044	&	0.19	&	0.12	&	0.04	&	0.13	&	0.10	&	0.11	\\
18173706$-$3405569	&	0.18	&	0.09	&	0.07	&	0.11	&	0.09	&	0.08	\\
18173994$-$3358331	&	0.16	&	0.11	&	0.00	&	0.12	&	0.11	&	0.12	\\
18182073$-$3353250	&	0.15	&	0.10	&	0.09	&	0.12	&	0.14	&	0.09	\\
18174478$-$3343290	&	0.14	&	0.07	&	0.14	&	0.09	&	0.06	&	0.08	\\
18183369$-$3352038	&	0.14	&	0.09	&	0.08	&	0.10	&	0.08	&	0.10	\\
18183098$-$3358070	&	0.12	&	0.09	&	0.04	&	0.09	&	0.10	&	0.09	\\
18180049$-$3246462	&	0.20	&	0.10	&	\nodata	&	\nodata	&	\nodata	&	\nodata	\\
18174067$-$3356000	&	0.15	&	0.08	&	0.03	&	0.09	&	0.10	&	0.09	\\
18173118$-$3358318	&	0.15	&	0.09	&	0.06	&	0.09	&	0.03	&	0.09	\\
18182052$-$3345251	&	0.17	&	0.09	&	0.05	&	0.08	&	0.02	&	0.09	\\
\enddata

\end{deluxetable}


\end{document}